\documentstyle[NATO,namedreferences,epsfig]{crckapb}
\newcommand{\apj}{Ap. J.}
\newcommand{\apjl}{Ap. J. Lett.}
\newcommand{\mnras}{M.N.R.A.S}
\newcommand{\etal}{{\it et al.}}

\newcommand{\npix}{{N_{\rm pix}}}
\newcommand{\transpose}{\raise0.1in\hbox{\kern-0.03in\scriptsize\ T}}
\newbox\grsign \setbox\grsign=\hbox{$>$} \newdimen\grdimen \grdimen=\ht\grsign
\newbox\simlessbox \newbox\simgreatbox
\setbox\simgreatbox=\hbox{\raise.5ex\hbox{$>$}\llap
     {\lower.5ex\hbox{$\sim$}}}\ht1=\grdimen\dp1=0pt
\setbox\simlessbox=\hbox{\raise.5ex\hbox{$<$}\llap
     {\lower.5ex\hbox{$\sim$}}}\ht2=\grdimen\dp2=0pt
\def\simgreat{\mathrel{\copy\simgreatbox}}
\def\simless{\mathrel{\copy\simlessbox}}

\newbox\propsign \setbox\propsign=\hbox{$\propto$}
\newdimen \propdimen \propdimen=\ht\propsign
\newbox\simproptobox
\setbox\simproptobox=\hbox{\raise.5ex\hbox{$\propto$}\llap
     {\lower.5ex\hbox{$\sim$}}}\ht3=\grdimen\dp3=0pt
\def\simpropto{\mathrel{\copy\simproptobox}}
\begin{opening}
\title{CALCULATION OF COSMIC BACKGROUND RADIATION 
ANISOTROPIES AND IMPLICATIONS}
\author{EMORY F. BUNN}
\institute{Astronomy Department \\
University of California, Berkeley}
\end{opening}

\begin{document}

\begin{abstract}
We review the physical processes that are thought to produce
anisotropy in the cosmic microwave background, 
focusing primarily (but not exclusively)
on the effects of acoustic waves in the
early Universe.  We attempt throughout to supply an intuitive, physical
picture of the key ideas and to elucidate the ways in which the
predicted anisotropy depends on cosmological parameters such as
$\Omega_0$ and $h$.
The second half of these lectures is devoted to a discussion
of microwave background data analysis techniques, with an emphasis on 
the analysis of the COBE DMR data.  In particular, the Karhunen-Lo\`eve
method of data compression is described in detail.
\end{abstract}

\section{Introduction}

Since the discovery four years ago of cosmic microwave background
(CMB) fluctuations (Smoot \etal~1992), the data from anisotropy
experiments have improved in both quality and quantity at a very rapid
pace.  CMB data already provide stringent constraints on cosmological
models, and with a plethora of balloon-borne and ground-based
experiments underway and two planned satellite missions, we can expect
further
dramatic improvement over the next decade.  In fact, there is a 
very real possibility that we will accurately measure many of the
most important cosmological parameters via the CMB anisotropy
spectrum (Jungman \etal~1996, Kosowsky \etal~1996).

In order to realize this promise, we must take great care in
developing tools for comparing observational data with theoretical
predictions.  Even with existing data, this process is far
from trivial, and with the much larger data sets of the near future
the task will become trickier.  There are at least two independent
problems to be faced: we must be able to make accurate predictions
of the anisotropy spectrum for any particular theory, and we
must develop adequate statistical techniques to facilitate
the comparison of these predictions with observations.\footnote{
Not to mention the far more difficult task of actually
gathering the data!}

These lectures are concerned with these two subjects.  We will first
review the primary physical mechanisms that are thought to be
responsible for generating CMB anisotropies.  The emphasis in this
half of the lectures will be on building an intuitive picture of the
relevant physical effects.  We will therefore give ourselves free
rein to make physically motivated approximations, rather than trying
to treat the rather involved subject of anisotropy formation with
complete precision.  This section of the lectures will draw heavily on
the work of Wayne Hu and Naoshi Sugiyama (Hu \& Sugiyama 1994, 1995a,
1995b, 1996; Hu 1995), as well as on a review article by Hu, Sugiyama,
\& Silk (1996) and two previous summer-school proceedings on the
subject (Hu 1996, Tegmark 1996c).

The second half of these lectures is devoted to issues of
statistics and data analysis.  We will study various ways in which
theoretical predictions of CMB anisotropy may be compared with data
sets.  Our primary focus will be on methods for analyzing the COBE DMR
data, since this is the largest and most powerful CMB data set in
existence; however, many of the issues that arise in analyzing the
COBE data are directly relevant to analyses of other experiments, both
present and future.  For example, we will pay special attention to the
issue of {\em data compression}; this subject was fairly important in
analyzing the COBE data, and its importance will only increase as CMB
data sets get larger and larger.  In particular, the planned MAP and
COBRAS/SAMBA missions will both return data sets several orders of
magnitude larger than COBE, and their analysis will therefore require
extensive data compression.

These lectures are organized as follows.
Section \ref{sec:overview} provides an overview of the key physical
processes that produce CMB anisotropy.  Section \ref{sec:primary}
discusses the primary anisotropy, including the Sachs-Wolfe effect
(Sachs \& Wolfe 1967) and anisotropies produced by acoustic
oscillations of the photon-baryon fluid (Peebles \& Yu 1970;
Doroshkevich, Zel'dovich,
\& Sunyaev 1978; Bond \& Efstathiou 1984), as well as
the diffusive damping of fluctuations (Silk 1968).  In
Section \ref{sec:secondary} we discuss anisotropies produced after
last scattering, such as the integrated Sachs-Wolfe effect (Sachs \&
Wolfe 1967, Rees \& Sciama 1968), the effect of gravitational
lensing (Blandford \& Narayan 1992, Seljak 1996b), 
and reionization (Sunyaev 1977, Silk 1982).  
Section \ref{sec:summary1}
attempts to synthesize the main ideas of the previous sections
and concludes the first half of these lectures.

The second half, which concerns issues of statistics and data
analysis, begins with Section \ref{sec:stats}, in which
we establish some basic results and notation having to do with
Gaussian random processes on the sphere.  Section \ref{sec:cmbdata}
presents a series of idealized thought experiments designed
to introduce
some of the key issues of CMB data
analysis.  This section also contains a digression on Bayesian
and frequentist statistical techniques.  In Section \ref{sec:cobe},
we apply what we have learned to an analysis of the four-year
COBE DMR data, and Section \ref{sec:summary} contains some
brief concluding remarks.

\section{An Overview of Anisotropy Formation}
\label{sec:overview}

CMB anisotropies encode large amounts of information about the
Universe.  Physical processes around the redshift of last scattering
(typically $z\simeq 1100$) produce the {\em primary anisotropy}, which
can be significantly altered by {\em secondary} processes between the
last-scattering surface and the present.  In addition, the angular
scale subtended by a particular source of anisotropy depends on the
spatial geometry as well as the distance to the last-scattering
surface.

With the exception of some effects at very low redshift, and
ignoring topological defect models, calculations
of CMB anisotropy are done in linear perturbation theory.
All of the relevant quantities are small perturbations
about a homogeneous Friedmann-Robertson-Walker solution.  Nonetheless,
making accurate numerical predictions of the CMB anisotropy in a
particular theory is a daunting numerical task.  In
a typical cold dark matter (CDM) model, the variables
one must keep track of include

\begin{itemize}
\item[$\bullet$] $\delta_{\rm B}\equiv \delta\rho_{\rm B}/\rho_{\rm B}$,
the baryon density perturbation.
\item[$\bullet$] $\delta_{\rm CDM}\equiv\delta\rho_{\rm CDM}/\rho_{\rm CDM}$, the
perturbation in the CDM density.
\item[$\bullet$] ${\bf v}_{\rm B}$, the baryon peculiar velocity field.
\item[$\bullet$] ${\bf v}_{\rm CDM}$, the CDM peculiar velocity field.
\item[$\bullet$] $\Psi$, essentially the Newtonian gravitational potential.
\item[$\bullet$] $\Phi$, the perturbation to the spatial curvature.\footnote{
We will work throughout in Newtonian gauge.  For our purposes
$\Psi$ and $\Phi$ are the only important perturbations to the
metric.  $\Psi$ is related to the perturbation to the time-time
component $g_{00}$ of the metric, and $\Phi$ has to do with the
perturbation to the spatial part $g_{ij}$.  For more information
on gauges, see the contribution of J.-L. Sanz to this volume, and
also Hu (1995, 1996) and references therein.}
\item[$\bullet$] $f_\gamma$, the photon phase-space distribution function.
\item[$\bullet$] $f_\nu$, the neutrino phase-space distribution function.
\end{itemize}

All of these quantities depend on position $\bf x$ and time $t$, and
$f_\gamma$ and $f_\nu$ are also momentum-dependent.  Their evolution
is governed by a nasty set of coupled partial differential equations.  For
the nonrelativistic species, we must keep track of the usual equations
of perturbation theory, namely the continuity equation, the Euler
equation, and the Poisson equation.  For the CDM, these equations
look like
\begin{eqnarray}
\label{eq:cont}
\dot\delta_{\rm CDM}+\nabla\cdot{\bf v}_{\rm CDM}&=&0, \\
\label{eq:euler}
\dot{\bf v}_{\rm CDM}+2{\dot a\over a}{\bf v}_{\rm CDM}&=&
-{1\over a^2}\nabla\Psi, \\
\label{eq:poisson}
\nabla^2\Psi&=&4\pi G\bar\rho\delta.
\end{eqnarray}
Here 
$a$ is the scale factor, $\bar\rho$ is the average
density, and a dot denotes a time derivative.  All spatial derivatives
are taken with respect to comoving coordinates.
In the last equation,
$\delta$ represents the total density perturbation, although we will
generally consider models that are gravitationally dominated by
CDM, so that we can replace $\delta$ with $\delta_{\rm CDM}$.
There are also continuity and Euler equations for the baryons,
the latter containing a pressure term.

The relativistic species (photons and neutrinos) are not characterized
by a simple velocity field, but by a distribution function whose
evolution is governed by the Boltzmann equation,
\begin{equation}
{Df\over Dt}\equiv {\partial f\over\partial t}+{\partial f\over\partial x^i}
{dx^i\over dt}+{\partial f\over\partial p}{dp\over dt}+
{\partial f\over\partial\gamma^i}{d\gamma^i\over dt}=C[f].
\end{equation}
Here $p$ is the magnitude of the momentum, $\gamma^i$ is
a direction cosine of the momentum, and $C$ is a collision term having
to do with scattering.  This equation applies to both $f_\gamma$
and $f_\nu$, although
at the epochs we are interested in the
neutrino collision term is zero.

In order to make accurate predictions of the CMB anisotropy
in a particular model, it is necessary to solve this system of equations
numerically.  If we work in Fourier space, we find that different
fluctuation modes are uncoupled and the solution is therefore
greatly simplified.  We write
\begin{equation}
\delta({\bf x},t)=\sum_{\bf k}\delta_{\bf k}(t)\exp(i{\bf k}\cdot{\bf x}),
\end{equation}
and similarly for the other quantities.  [For the distribution
functions, it is convenient to make a second expansion in Legendre
polynomials $P_l(\hat {\bf k}\cdot\hat{\bf p})$.]  
The fact that different $\bf
k$-modes decouple makes the problem computationally tractable.
Furthermore, as we shall see, the fact that we can work with one mode
at a time makes it easier to get a conceptual understanding of
anisotropy formation.

In recent years excellent codes have been developed for integrating
these equations.  [See Hu \etal~(1995) and Bond (1996)
for fairly recent
discussions of the state of the art, and Seljak \& Zaldarriaga
(1996) for an important
subsequent development.]  We will not discuss the details of such
precise calculations here; rather, we will follow a less precise but
more intuitive picture of the formation of anisotropies, based on a
series of physically motivated approximations.  This approach
makes it easier to see what the important physical processes are
and also gives us an understanding of how various features in
the anisotropy spectrum depend on key cosmological parameters.

We will begin by discussing the sources of primary anisotropy: the
Sachs-Wolfe effect (Sachs \& Wolfe~1967), which describes
gravitational red- and blueshifts due to potential differences on the
surface of last scattering;
the Doppler effect due to bulk motions of
the last-scattering surface (Sunyaev \& Zel'dovich 1970); 
and intrinsic temperature variations from
point to point (Silk 1967).  We will then discuss some sources of secondary
anisotropy, the most important of which is the integrated Sachs-Wolfe
(ISW) effect, which describes energy changes in photons as they pass
through time-varying potentials.  [This effect was also
treated by Sachs \& Wolfe (1967), as well as by Rees \& Sciama (1968)
at nearly the same time.]
Other secondary sources of
anisotropy include scattering by reionized matter and gravitational
lensing.

\begin{figure}[t]
\centerline{\epsfig{file=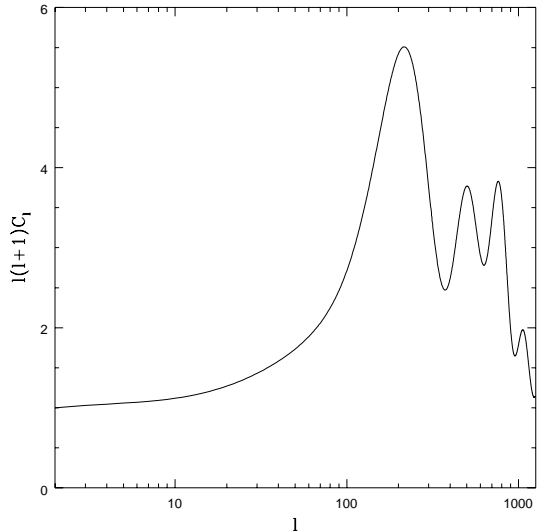,width=3in}}
\caption{The angular power spectrum $l(l+1)C_l$ for a standard cold
dark matter model.  The parameters of this model are as follows:
$n=1$, $h=0.5$, $\Omega_0=1$, $\Omega_{\rm B}h^2=0.013$.  This power spectrum
was computed by N. Sugiyama.}
\label{fig:scdm}
\end{figure}

At first, we will consider the evolution of only 
one Fourier mode at a time; however, we will eventually
need to synthesize all of the different Fourier modes together
to see what the total CMB anisotropy on the sky looks like.
To do that, we will need to know the {\em power spectrum}
of the density perturbation.  This is simply the mean-square
amplitude of the various Fourier modes:
\begin{equation}
P(k)=\langle|\delta_{\bf k}|^2\rangle.
\end{equation}
(As long as space is isotropic, $P$ depends only on the magnitude
of $\bf k$.)  The angle brackets here denote an ensemble average,
although it is frequently acceptable to assume $\delta$ is ergodic,
in which case the angle brackets can equally well be regarded
as a spatial average.\footnote{
Beware: When we describe $\Delta T/T$ as a random field on the sphere,
we may {\em not} assume ergodicity: $\Delta T/T$ is
never ergodic.}
We often assume that the initial power spectrum
is a power law in $k$: $P(k)\propto k^n$.  As we will see
below, the analogous quantity for describing the observed CMB
anisotropy is the {\em angular power spectrum}:
\begin{equation}
C_l=\langle|a_{lm}|^2\rangle.
\end{equation}
Here $a_{lm}$ is a coefficient of an expansion of a spherical
harmonic expansion of the temperature anisotropy (spherical
harmonic expansions being the natural analogue of Fourier expansions
for data sets that live on the sphere).  A mode with spherical harmonic
index $l$ probes an angular scale on the sky of $\theta\sim l^{-1}$.
In any particular
cosmological model, the angular power spectrum
$C_l$ is related linearly to the matter power spectrum $P(k)$.
The angular power spectrum for a CDM model is shown in Figure 
\ref{fig:scdm}.\footnote{
The prefactor $l(l+1)$ in Figure \ref{fig:scdm} (and all of the
other power spectrum plots we will see)
is traditional.  In a flat cosmological model
with an $n=1$ power spectrum, the Sachs-Wolfe contribution to the
power spectrum is proportional to $1/l(l+1)$.  The Sachs-Wolfe effect
dominates on large scales, explaining the flatness of Figure \ref{fig:scdm}
at low $l$.  The quantity $l(l+1)C_l$ is also approximately proportional
to the total power per logarithmic interval in $l$.  (To make
this proportionality exact, one would use $l(l+{1\over 2})C_l$ instead.)}
The primary goal of Section \ref{sec:primary} will be to explain
the multiple peaks in this spectrum.

\section{Primary Anisotropies}
\label{sec:primary}

\subsection{The Gravitational Potential}

We will begin by assuming that, after the end of 
the radiation epoch, most of the mass in the Universe
is in the form of cold dark matter:
\begin{equation}
\Omega_{\rm B}\ll\Omega_{\rm CDM}.
\end{equation}
Then the gravitational potential is completely determined by the CDM,
and three equations
(\ref{eq:cont}$-$\ref{eq:poisson}) can be solved for $\Psi$ and $\delta$
without worrying about what the other species are doing.  Then,
once we know the gravitational potential $\Psi$, we can solve for 
the evolution of the photons and baryons.

Equations (\ref{eq:cont}$-$\ref{eq:poisson}) can be combined
into a single second-order equation for $\delta$,
\begin{equation}
\ddot\delta+2{\dot a\over a}\dot\delta-4\pi G\bar\rho\delta=0.
\end{equation}
At early times, when the Universe is radiation dominated, the
last term in this equation is negligible, and the two
linearly independent solutions
are $\delta=\hbox{const.}$ and $\delta\propto\ln t$.  There
is therefore little growth during the radiation era.

If the Universe is matter dominated (meaning that both 
radiation and curvature are negligible in the Friedmann equation), then
we have $a\propto t^{2/3}$, and the solutions are $\delta\propto t^{2/3}
\propto a$ and $\delta\propto t^{-1}$.  At late times, of course,
the growing mode is the one that matters.
If we plug the matter-dominated growing-mode solution into
the Poisson equation (\ref{eq:poisson}), we find that {\em $\Psi$ is
independent of time}.  This is a key fact, to which we will return
repeatedly.

\subsection{The Photon-Baryon Fluid}

Now that we know what the gravitational potential is doing, we are
ready to study the evolution of the photons and baryons.  We
do this by making another approximation: we assume {\em tight coupling}
between photons and baryons.  Specifically, we assume that the
mean free time $\tau$ between photon collisions is small compared
to the other important time scales:
\begin{equation}
\tau \ll H^{-1}, (ck)^{-1}, (c_sk)^{-1}.
\end{equation}
Here $H^{-1}$ is the expansion time scale, $(ck)^{-1}$ is the
light-travel time across a Fourier mode, and $(c_sk)^{-1}$ is the
sound-travel time across a mode ($c_s$ being the sound speed).
This is an excellent approximation right up until around the
time of last scattering.

In the tight-coupling approximation, frequent scattering isotropizes
the photon distribution function $f_\gamma$: at any particular
point, $f_\gamma$ is isotropic in the rest frame of the baryons
at that point.  In fact, $f_\gamma$ is completely characterized
by the temperature distribution.  Furthermore, the photon
and baryon densities are coupled adiabatically: $n_\gamma\propto
n_{\rm B}\propto T^3$.  The behavior of the photon-baryon fluid
is therefore characterized by a single variable: if we know,
say, $\delta_{\rm B}({\bf x},t)$, we can determine $\bf v_{\rm B}$, $T$,
and $f_\gamma$.  We will find it convenient to take as our
variable the fractional temperature fluctuation, which is simply one third
of the baryon density fluctuation:
\begin{equation}
\Theta({\bf x},t)\equiv {\Delta T\over T}({\bf x},t)={1\over 3}\delta({\bf
x},t).
\end{equation}

With these approximations, the dynamics of the photon-baryon fluid
is described by the single equation
\begin{equation}
{d\over d\eta}\left[(1+R)\dot\Theta\right]+{k^2\over 3}\Theta=F(\eta).
\label{eq:master}
\end{equation}
This equation comes from the Euler and continuity equations for the
fluid.  We are working in units in which $c=1$.
For a derivation of this equation, see Hu (1995). In this equation, $\eta$
is the conformal time,
\begin{equation}
\eta=\int^t {dt\over a(t)},
\end{equation}
and $R\equiv 3\rho_{\rm B}/4\rho_\gamma$ is essentially the baryon-to-photon
energy ratio.  The overdot denotes a derivative with respect to
conformal time.
This equation is in Fourier space, so $\Theta=\Theta_{\bf k}$
represents a single Fourier mode with wavenumber $\bf k$.\footnote{
It has become standard practice in cosmology
to denote functions and their Fourier
transforms by the same symbol, relying on context to tell the difference.
[For the only recent exception I know about, see Tegmark (1996c).]
Odious as this practice is, I have bowed to convention in these lectures.}
The right-hand side $F(\eta)$ is a gravitational driving term,
\begin{equation}
F(\eta)=-{k^2\over 3}(1+R)\Psi-{d\over d\eta}\left[(1+R)\dot\Phi\right].
\end{equation}

The rest of this section will be devoted almost entirely to a discussion
of the solution of equation~(\ref{eq:master}).  We begin by making
some useful observations.  First,
\begin{equation}
R=\left(450\over 1+z\right)\left(\Omega_{\rm B}h^2\over 0.015\right),
\end{equation}
where $h$ is the Hubble parameter in units of $100\,\rm km\,s^{-1}\,
Mpc^{-1}$, $z$ is the redshift, and $\Omega_{\rm B}$ is the baryonic
contribution to the density parameter.  So for standard recombination
at $z\simeq 1000$ and baryon densities around the nucleosynthesis
range, $R\simeq {1\over 2}$ at the time of last scattering.

With the approximations that we're making, there are no
anisotropic stresses, so the two gravitational
potentials are simply related to each other:
\begin{equation}
\Phi=-\Psi.
\end{equation}
Furthermore, we have seen that during the matter-dominated epoch,
if linear theory is valid, $\Psi$ is independent of time.  The
gravitational driving term therefore simplifies to
\begin{equation}
F(\eta)=-{k^2\over 3}(1+R)\Psi.
\end{equation}

\subsection{Acoustic Oscillations}

To develop an intuitive feel for the solutions to equation (\ref{eq:master}),
we will start by making some excessive and unwarranted approximations.
We will then gradually relax those approximations to get a more accurate
picture.  First, let's assume that $R$ and $\Psi$ are independent of time.
Then
\begin{equation}
(1+R)\ddot\Theta+{k^2\over 3}\Theta=-{k^2\over 3}(1+R)\Psi.
\end{equation}
This is the equation for a simple harmonic oscillator, with solution
\begin{equation}
\Theta(\eta)=-(1+R)\Psi+K_1\cos(kc_s\eta)+K_2\sin(kc_s\eta).
\end{equation}
Here $K_1$ and $K_2$ are constants to be fixed by the initial
conditions and $c_s=(3(1+R))^{-1/2}$ is the sound speed.
In this approximation, then, each Fourier mode represents 
an acoustic plane wave propagating at speed $c_s$.

There is a simple physical picture underlying this result.  The
baryon-photon fluid wants to fall into the potential wells, but
it is supported by radiation pressure.  The balance
between pressure and gravity sets up acoustic
oscillations.  The three terms in equation
(\ref{eq:master}) come from the inertia of the fluid, the
radiation pressure, and the gravitational field.

In fact, let's make things even simpler and set $R=0$.  Then
\begin{equation}
\Theta(\eta)=-\Psi+K_1\cos(kc_s\eta)+K_2\sin(kc_s\eta).
\end{equation}
In many theories, the initial perturbation is {\em adiabatic},
meaning that the matter and radiation fluctuations are the
same at any particular point.  With these initial conditions,
$\dot\Theta=0$ at very early times, and $\Theta(0)=-2\Psi/3$,
so
\begin{equation}
\Theta(\eta)=-\Psi+\hbox{$1\over3$}\Psi\cos kc_s\eta.
\label{eq:simplestsol}
\end{equation}

Continuing to focus our attention on a single Fourier mode, let
us determine what kind of anisotropy we would expect to see
on the sky.  As we have mentioned, the three sources of primary
anisotropy are gravity, the Doppler effect, and intrinsic temperature
variations,
\begin{equation}
{\Delta T\over T}=\left[\Psi+\hat{\bf r}\cdot{\bf v}+\Theta\right]_{\eta
=\eta_{\rm LS}},
\label{eq:primary}
\end{equation}
where $\eta_{\rm LS}$ is the time of last scattering and $\hat{\bf r}$ is
a unit vector in the direction of observation.

Ignoring the Doppler term for the moment, note that the other
two terms give a pure cosine oscillation,
\begin{equation}
\Psi+\Theta=\hbox{$1\over 3$}\Psi\cos kc_s\eta,
\label{eq:thetapsi}
\end{equation}
so the r.m.s. $\Delta T/T$ is large when $kc_s\eta_{\rm LS}$ is
an integer multiple of $\pi$.  Therefore, if the initial conditions
have a smooth power spectrum, $\Delta T/T$ will have a harmonic
series of peaks in $k$-space, leading to a harmonic series in the
angular power spectrum of anisotropy on the sky.  This is the
origin of the so-called ``Doppler peaks'' in Figure \ref{fig:scdm}.
Ironically, the peaks have nothing to do
with the Doppler effect.  In fact, the peaks are caused
by modes that have reached maxima of compression and rarefaction
at the time of last scattering; the Doppler contribution to the
anisotropy in these modes is zero!

The first peak is caused by modes that have had time to
oscillate through exactly one half of a period before last scattering;
the modes that cause the second peak have oscillated through
a full period, and so on.
The physical scale of the first peak is therefore $\lambda\sim k^{-1}=c_s
\eta_{\rm LS}/\pi\sim 30\,\rm Mpc.$  The distance to the last-scattering
surface is $D\equiv\eta_0-\eta_{\rm LS}\sim 6000\,\rm Mpc$, 
so the angular scale of the first
peak is $\lambda/D\sim 0^\circ\!\!.25$.  We will be more precise about
the correspondence between physical scales and angular scales
later.

Earlier, we threw out the Doppler term in equation (\ref{eq:primary})
for no particular reason.  We had better put it back.  Using the
continuity equation (\ref{eq:cont}) and the relation $\delta=3\Theta$,
we find that 
\begin{equation}
{\bf v}={3i\over k}\dot\Theta\hat{\bf k}.
\end{equation}
Here $\hat{\bf k}$ is a unit vector in the direction of ${\bf k}$
and ${\bf v}$ and $\delta$ are still in Fourier space.  
Differentiating equation (\ref{eq:simplestsol}) and using the
fact that $c_s=1/\sqrt 3$ for $R=0$, we find that 
$\dot\Theta=-{1\over 3\sqrt 3}\Psi\sin kc_s\eta$.  Since the r.m.s.
value of $\hat{\bf r}\cdot \hat{\bf k}$ is $1/\sqrt 3$,
the r.m.s. Doppler contribution to equation (\ref{eq:primary})
is 
\begin{equation}
\left[\Delta T\over T\right]_{\hbox{\scriptsize Doppler}}=
{i\over 3}\Psi\sin kc_s\eta.
\label{eq:doppler}
\end{equation}
This has the same amplitude as the $(\Theta+\Psi)$ contribution,
but is $90^\circ$ out of phase in both time (it goes like a sine
instead of a cosine) and space (it has an extra factor of $i$).  
This has the rather disastrous
consequence of completely erasing the Doppler peaks: the total
$\Delta T/T$ is the quadrature sum of (\ref{eq:thetapsi}) and
(\ref{eq:doppler}):
\begin{equation}
\left(\Delta T\over  T\right)^2\propto\sin^2 kc_s\eta+
\cos^2kc_s\eta=1.
\end{equation}
The $k$ dependence, which led to the peaks, is gone.

\begin{figure}[t]
\centerline{\epsfig{file=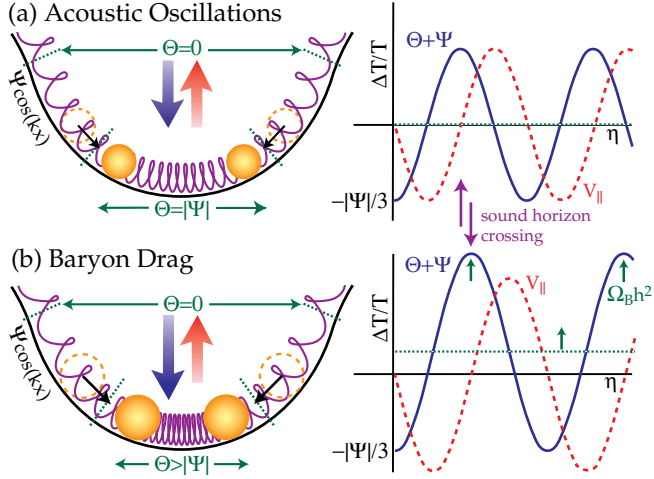,width=3.5in,angle=-90}}
\caption{
A simple mechanical model for a single mode of
acoustic oscillation of the photon-baryon
fluid.  The behavior of the fluid inside of a potential well is shown;
the behavior atop a potential hill would be the reverse.  The
springs represent the restoring force of the photon pressure and
the balls represent the effective mass of the system.  The top
panel shows the case where the baryon contribution to the effective
mass can be neglected, and the lower panel shows the effect of including
baryons.  Baryons increase the mass of the fluid, causing a displacement
of the zero point of the oscillations.  In addition, the sound
speed is lowered.  This has two effects, both of which may be
seen in the plots on the right: baryons make
the oscillations proceed more slowly
and also reduce the Doppler contribution to $\Delta T/T$ relative
to the intrinsic and Sachs-Wolfe contributions.  Reprinted from
Hu (1996).
}
\label{fig:cartoonbaryon}
\end{figure}

The problem, of course, is that we have taken our approximations too far.
Specifically, the culprit is the limit $R\to 0$.  Physically,
taking the limit $R\to 0$ means ignoring the dynamical effects of
the baryons.
Let us remove
that assumption, but keep the approximation that $R$ is time-independent.
Then the solution for $\Theta(\eta)$ changes in two ways.  
The sound speed gets smaller by a factor $1/\sqrt{1+R}$, and the
driving term $F(\eta)$ gets bigger by a factor $1+R$.
The adiabatic solution to equation (\ref{eq:master}) is now
\begin{equation}
\Theta(\eta)=\hbox{$1\over 3$}(1+3R)\Psi\cos kc_s\eta
-(1+R)\Psi.
\end{equation}

By allowing $R$ to be nonzero, we have increased the amplitude
of the cosine oscillations by a factor $(1+3R)$.  Furthermore,
there is now an offset in the combined Sachs-Wolfe and adiabatic
contributions to $\Delta T/T$: in the limit $R\to 0$, we found
that $\Theta+\Psi$ oscillated symmetrically about zero; now
it oscillates about $-R\Psi$.  Most important, a nonzero $R$
reduces the amplitude of the Doppler contribution to the
anisotropy, relative to the Sachs-Wolfe contribution, since
$v$ is proportional to $c_s\Theta$ and $c_s$ has gotten smaller.
Since the cosine oscillations are now larger in amplitude than
the sine oscillations, we do indeed expect to see a 
series of peaks at $kc_s\eta_{\rm LS}=m\pi$.

Why does including the dynamical effect of the baryons effect
these changes in the solution?  The essential reason is that
baryons contribute to the {\em effective mass} of the photon-baryon fluid,
but not to the {\em pressure}.  (This is clear from looking at equation
(\ref{eq:master}): the first term, representing the effective mass,
depends on $R$, but the second term, representing pressure support, does
not.)  The effect of the baryons, therefore, is to slow down
the oscillations, and also to make the fluid fall deeper into
the potential wells.  This explains all three of the key effects
we have just mentioned: the increased oscillation amplitude,
the offset in the center of the oscillations, and the reduction
in importance of the velocity term relative to the other terms.
These effects are represented pictorially in Figure \ref{fig:cartoonbaryon}.

\begin{figure}[t]
\centerline{\epsfig{file=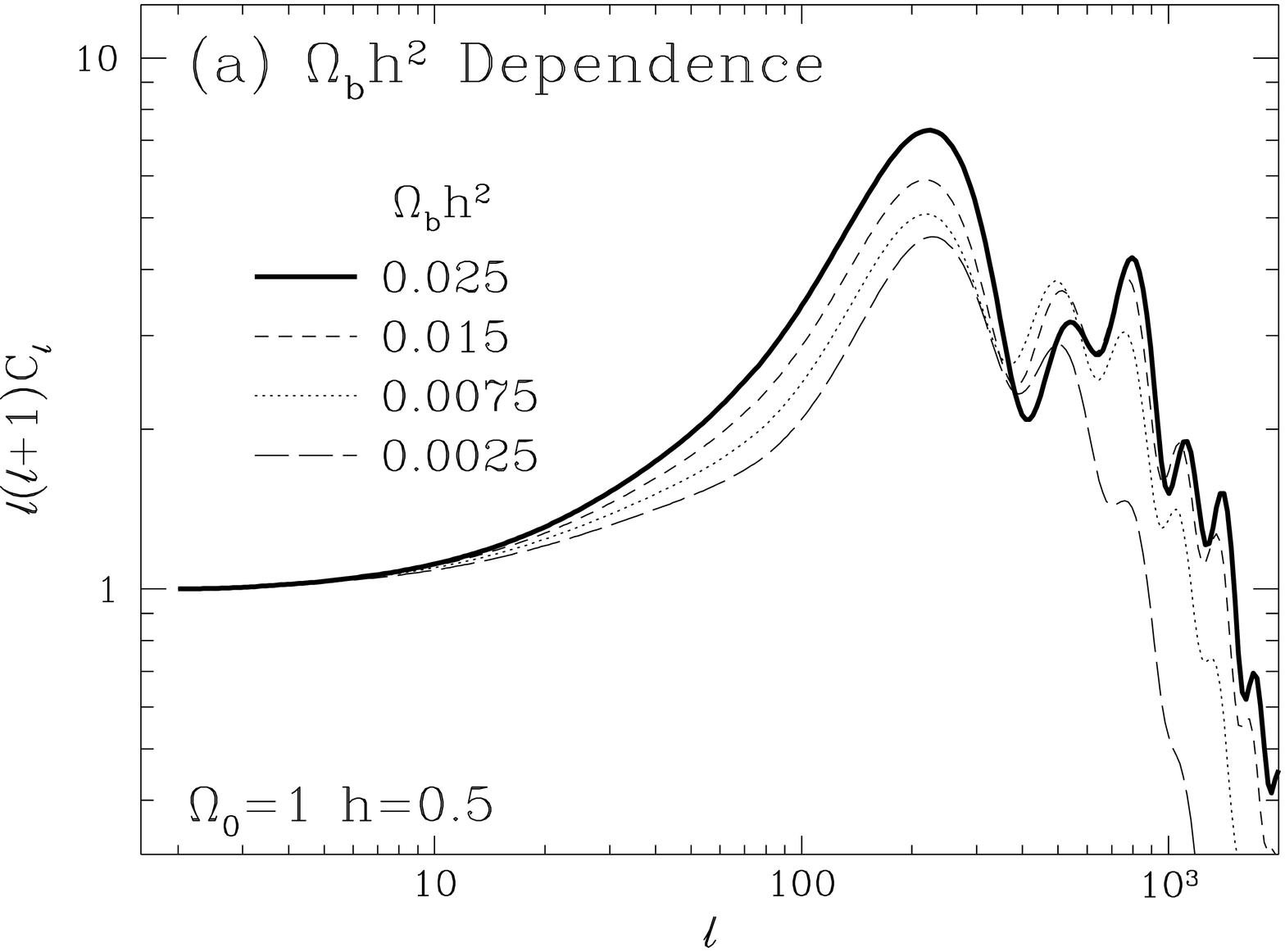,width=3in}}
\caption{Angular power spectra for CDM models with varying values of
the baryon density $\Omega_{\rm B}h^2$.  Reprinted from Hu (1996).}
\label{fig:clbaryon}
\end{figure}

Based on this analysis, we can predict that the height of the peaks
in the CMB anisotropy spectrum should depend on the baryon density:
the larger the baryon density, the larger $R$, and the greater the
amplitude of the oscillations.  Furthermore, because of the offset
in the oscillations, we expect the odd-numbered peaks to be enhanced
relative to the even-numbered ones.  (In the language
of Figure \ref{fig:cartoonbaryon}, the compressions produce
larger anisotropies than the rarefactions.  Of course, if we had
chosen to draw a potential peak instead of a potential well
in Figure \ref{fig:cartoonbaryon}, we would make precisely
the opposite statement.)

Both of these effects are
found in detailed calculations and can be 
seen in Figure \ref{fig:clbaryon}.

We can make further refinements to these approximations without
too much difficulty.  For instance, we can allow $R$ to vary
with time.  The time scale on which $R$ varies is of order a
Hubble time and is much longer than the period of the acoustic
waves.  We can therefore treat the variation of $R$ (and the
concomitant variation in $c_s$) in the WKB approximation.
There are two main results.  First, the phase of the oscillation
changes from $kc_s\eta$ to $k\int c_s\,d\eta$.  Second, the
amplitude of the oscillations grows with time in proportion to
$c_s^{1/2}$, or $(1+R)^{-1/4}$.\footnote{The easiest way to see this
is to note that $m\omega A^2$ is an adiabatic invariant for
a harmonic oscillator.  Here $m$ is the mass, $A$ is the
amplitude, and $\omega$ is the frequency.  Of course, the
result can also be derived directly from the WKB approximation.}

\subsection{Driving}
\label{sec:driving}

\begin{figure}[t]
\centerline{\epsfig{file=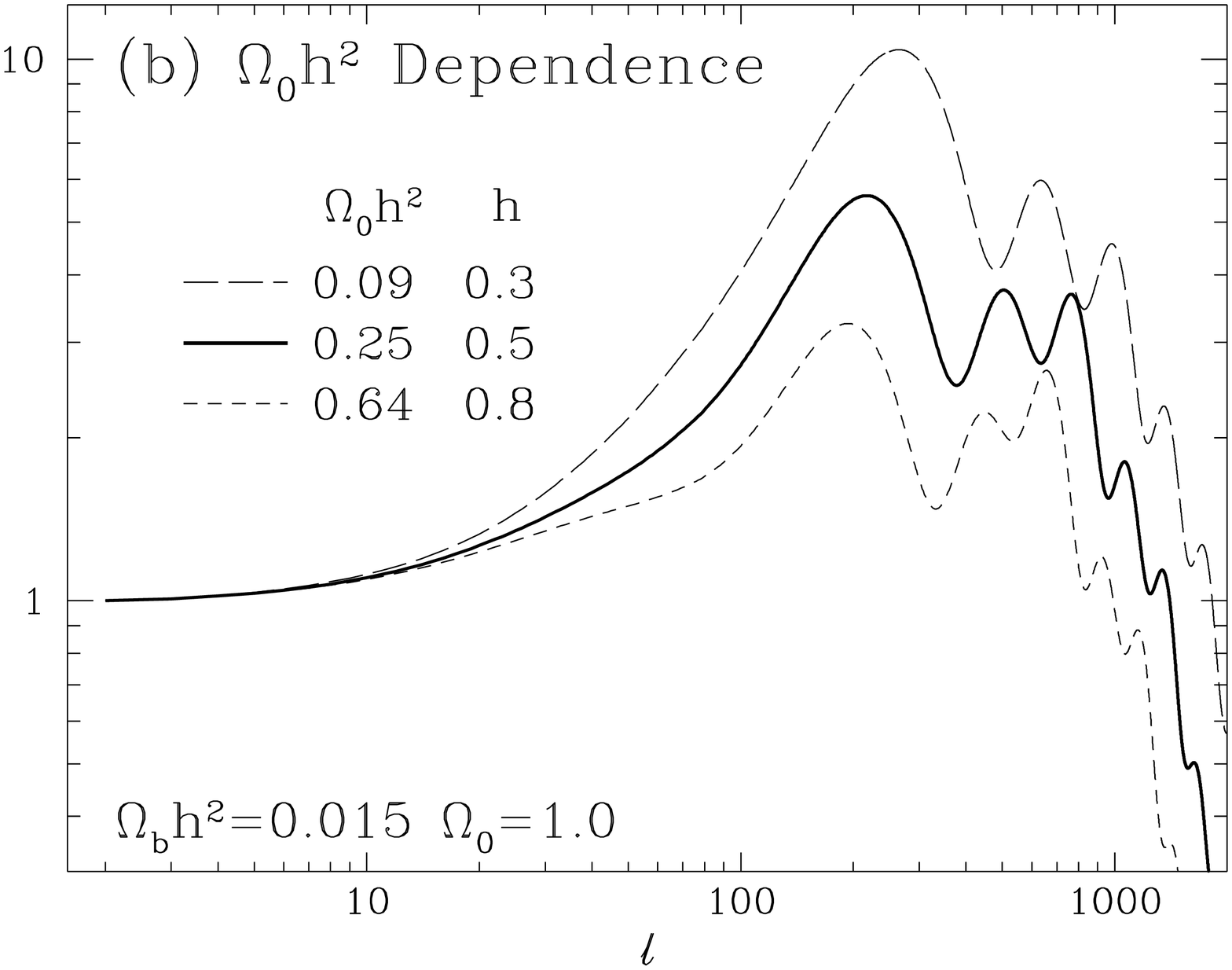,width=3in}}
\caption{Angular power spectra for CDM models with varying values
of $h$.  All of these models have $\Omega_0=1$.  For lower values
of $\Omega_0h^2$, matter domination occurs later.
The driving effect of the decay in the gravitational
potential is therefore more significant, increasing the peak height.
Reprinted from Hu (1996).}
\label{fig:clh}
\end{figure}

We can also relax the approximation that $F(\eta)$ is constant
in time.  This has interesting consequences.  A constant term
on the right-hand side of an oscillator equation merely offsets
the center of the oscillations; in contrast, a time-varying term
genuinely drives oscillations.  In particular, if the driving term
varies significantly on a time scale comparable to the period
of the oscillations, resonant driving can occur.

We have seen that $\Psi$ (and hence $F$) is constant during matter
domination, but it decays during the radiation epoch.  For modes
that enter the horizon before matter domination, $\Psi$ decays
while that mode is undergoing its oscillations.  The decay in $\Psi$
therefore boosts the amplitude of those short-wavelength modes.
The modes that receive the largest boost are those that entered
the horizon before matter-radiation equality at a redshift
\begin{equation}
z_{\rm eq}=24000\Omega_0h^2.
\end{equation}
These modes are characterized by wavenumbers
\begin{equation}
k\simgreat k_{\rm eq}=(14\,{\rm Mpc})^{-1}\Omega_0h^2.
\label{eq:keq}
\end{equation}

\begin{figure}[t]
\centerline{\epsfig{file=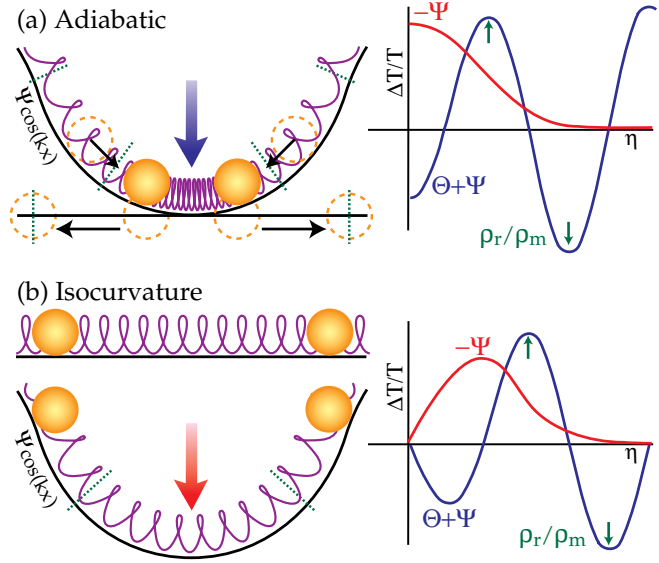,width=3.5in,angle=-90}}
\caption{
Driving effects on the acoustic oscillations.  (a) In the adiabatic
case, the gradual decay in the potential causes a relatively
small increase in the amplitude of the oscillations.  (b) For
isocurvature initial conditions, the initial perturbation $\Theta+\Psi$
is zero, and the growth (and subsequent decay) of $\Psi$ is entirely
responsible for driving the oscillations.  Reprinted from Hu, Sugiyama,
\& Silk (1996).}
\label{fig:cartooniso}
\end{figure}

The effect of the driving term becomes evident
if we look at power spectra for critical-density
models with different
values of the Hubble parameter: for low $h$, matter domination
occurs later and the boosting effect is greater.  This effect
is shown in Figure \ref{fig:clh}.

We have been focusing on models with adiabatic initial conditions.
If we instead consider {\em isocurvature} models, the effect
of the driving term becomes even more evident.  In isocurvature
models, the total density perturbation vanishes at
early times:
\begin{equation}
\delta\rho_{\hbox{\scriptsize total}}=\delta\rho_{\rm B}+\delta\rho_\gamma
+\delta\rho_{\rm CDM}+\ldots=0.
\end{equation}
Clearly $\Theta(0)=0$ in these models.  As time passes, $\delta\rho_\gamma$
redshifts away, leaving genuine density perturbations and hence
nonzero potentials $\Phi$ and $\Psi$.  Oscillations are therefore
driven in $\Theta$.  In contrast to the adiabatic
case, these isocurvature oscillations are proportional to $\sin
kc_s\eta$ rather than $\cos kc_s\eta$.  The peaks in an isocurvature
spectrum are therefore different in phase from adiabatic peaks.
The peak locations in the CMB anisotropy spectrum can 
distinguish quite robustly between adiabatic and isocurvature models.
Figure \ref{fig:cartooniso} illustrates the origin of the peaks
in isocurvature models.

\subsection{Damping}

\begin{figure}[t]
\centerline{\epsfig{file=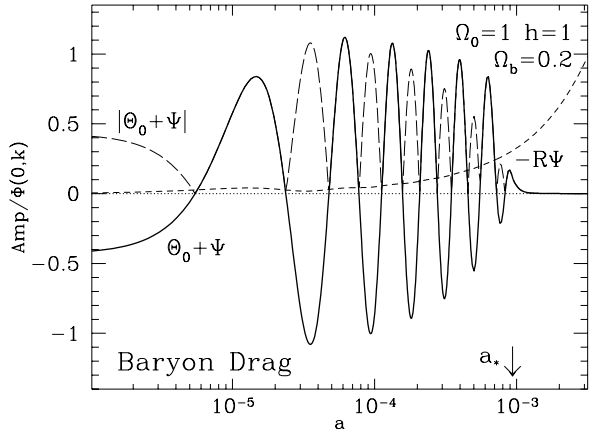,width=3in}}
\caption{The time evolution of a single Fourier mode.
$a$ is the scale factor, normalized to unity today.  $a_*$ is
the scale factor at recombination.  At early times potential decay
increases the amplitude of the oscillations.  The 
heights of the positive and negative peaks are offset by $-R\Psi$
with respect to each other.  The decline in amplitude at late
times is due to diffusion damping.  Reprinted from Hu (1996).
}
\label{fig:onemode}
\end{figure}

We have been assuming so far that the tight-coupling approximation
holds perfectly right up until the moment $\eta_{\rm LS}$, and
that the photons are instantaneously released at that moment.
In fact, the failure of the tight-coupling approximation, especially
around the time of last scattering, causes significant damping
of fluctuations as photons diffuse out of hot, overdense regions.
Furthermore, the last-scattering ``surface'' is really a shell
of some thickness.  Oscillations on scales smaller than
this thickness do not show up as observable anisotropies on the sky,
since any particular line of sight
will look at multiple peaks and troughs of that mode.

To get a rough estimate of the importance of diffusion damping (also
known as Silk damping), consider a photon undergoing a random
walk through the photon-baryon fluid.  If the mean free path
is $\lambda$, then at a time $\eta$, a typical photon has scattered
about $N\sim\eta/\lambda$ times and has diffused through a distance
$\lambda_{\rm D}\sim\sqrt{N}\lambda\sim\sqrt{\eta\lambda}$.
If a particular Fourier mode has a wavelength less than this
diffusion length, then the photons will have diffused from overdense
to underdense regions, and the mode will be damped away.
Diffusion damping thus occurs for modes with $k^{-1}\simgreat \lambda_{\rm D}$.
Most of the damping occurs around the time of last scattering, since
that is when the mean free path $\lambda$ becomes large.

In Figure \ref{fig:onemode} we show the time evolution of a particular
mode, including the damping at the end, and in Figure \ref{fig:decomp}
below
we show the net effect of diffusion damping on a CMB power spectrum.

\subsection{Projection}

In order to complete the story of primary anisotropies, we need
to specify precisely how a particular plane wave is projected
onto a specific angular scale on the sky.  It is clear that
a mode with wavelength $\lambda$ will show up on an angular
scale $\theta\sim\lambda/R$, where $R$ is the distance to
the last-scattering surface, or in other words, a mode with
wavenumber $k$ shows up at multipoles $l\simpropto k$.
Consequently, tilting the spectral index
$n$ of the primordial matter power spectrum essentially
just tilts the angular power spectrum.
Let us now make this rough observation
mathematically precise.

\begin{figure}[t]
\centerline{\epsfig{file=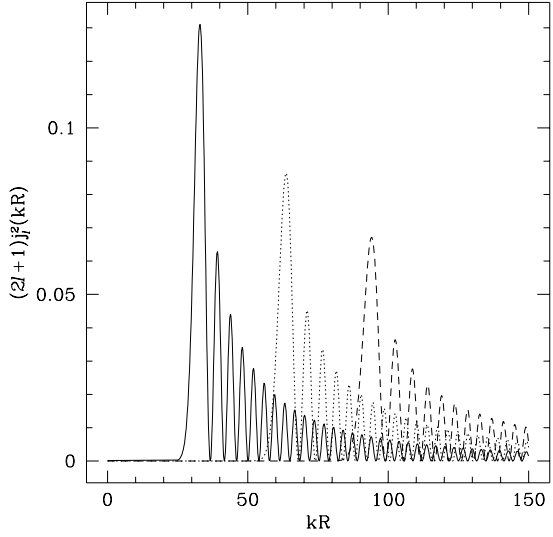,width=3in}}
\caption{The quantity $(2l+1)j_l^2(kR)$ is plotted for $l=30$,
$l=60$, and $l=90$.  This quantity determines how much power a Fourier
mode with wavenumber $k$ contributes to multipole $l$.  Note
that, while most of the power is deposited at $l\simeq kR$, there
is significant ``bleeding'' to lower $l$.}
\label{fig:bessj}
\end{figure}

If we are looking in a direction $\hat{\bf r}$ in the sky, then
(ignoring the thickness of the last-scattering surface), the
anisotropy we see is simply $\Delta T/T(\hat{\bf r})=\Theta^{\rm (tot)}
(R\hat{\bf r})$, where $\Theta^{\rm (tot)}$ includes all three
terms in equation (\ref{eq:primary}).  For a single Fourier mode, this
is simply
\begin{equation}
{\Delta T\over T}(\hat{\bf r})=\Theta_{\bf k}^{\rm (tot)} 
\exp(i{\bf k}\cdot\hat{\bf r}R).
\label{eq:onemode-real}
\end{equation}
To quantify the amount of power this produces on different angular scales,
we expand in spherical harmonics $Y_{lm}$.  The relevant identity is
(Jackson 1975)
\begin{equation}
\exp(ikR\hat{\bf k}\cdot\hat{\bf r})=4\pi\sum_{l,m}i^lj_l(kR)
Y_{lm}^*(\hat{\bf k})Y_{lm}(\hat{\bf r}).
\label{eq:ylmident}
\end{equation}
Combining equations (\ref{eq:onemode-real}) and (\ref{eq:ylmident}),
we find that
\begin{equation}
{\Delta T\over T}(\hat{\bf r})=\sum_{l,m}a_{lm}Y_{lm}(\hat{\bf r}),
\end{equation}
where
\begin{equation}
a_{lm}=4\pi\Theta_{\bf k}^{\rm (tot)}i^lj_l(kR)Y_{lm}^*(\hat{\bf k}).
\end{equation}
The total power produced by this mode in the multipole $l$ is
\begin{equation}
a_l^2\equiv\sum_{m=-l}^l |a_{lm}|^2=4\pi(2l+1)\left|
\Theta_{\bf k}^{\rm (tot)}\right|^2
j_l^2(kR).
\end{equation}
The spherical Bessel function $j_l(x)$ peaks at $x\sim l$, so a single
Fourier mode $\bf k$ does indeed contribute most of its power around
multipole $l_k=kR$, as expected.  However, as Figure \ref{fig:bessj}
shows, $j_l$ does have significant power beyond the first peak,
meaning that the power contributed by a Fourier mode ``bleeds''
to $l$-values lower than $l_k$.  This is due to the fact that 
a mode appears to have a longer wavelength when looked at along
a line of sight nearly perpendicular to the wavevector.

These formulae assume that the Universe is spatially flat.  If there
is curvature, then the correspondence between physical scales
at last scattering and angular scales on the sky changes.  In an
open Universe, for example, geodesics focus in such a way that
a particular angular scale corresponds to a much larger physical
scale on the last-scattering surface.  A particular Fourier
mode in an open Universe projects to multipoles $l\sim kR_A$,
where $R_A$ is the {\em angular-diameter distance} to the last-scattering
surface, given by
\begin{equation}
R_A={1\over\sqrt{|K|}}\sinh\left(\sqrt{|K|}R\right).
\end{equation}
Here $K$ is the curvature.  When $|K|$ is small, $R_A\to R$, but
for large $|K|$ (low $\Omega_0$), $R_A$ grows exponentially with 
metric distance.

\begin{figure}[t]
\centerline{\epsfig{file=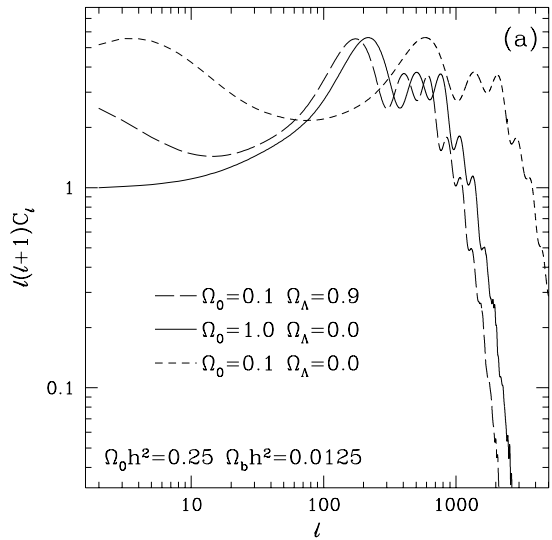,width=3in}}
\caption{$\Omega_0$-dependence of the angular power spectrum.
In open models, the angular-diameter distance to the last-scattering
surface is large, so the features in the power spectrum are shifted
to small angular scales.  In a flat model with a cosmological constant,
the distance to the last-scattering surface is larger than in
an $\Omega_0=1$ model, but the size of
the sound horizon also increases, producing little net effect on the
location of the peaks.  The structure at low $l$ in the low-density
models is due to the integrated Sachs-Wolfe effect.  Reprinted from
Hu \& White (1996).
}
\label{fig:clopen}
\end{figure}

This projection effect is easy to see in predictions of the CMB anisotropy.
In an open Universe, features such as the acoustic peaks and the
damping scale are shifted towards smaller angular scales, {\it i.e.},
towards higher $l$.  (See Figure \ref{fig:clopen}.)

Note that the approximate linear relation between $l$ and $k$ holds
only for primary anisotropies.  The secondary anisotropies, which
we discuss below, tend to occur at a wide range of distances (in
contrast to
the relatively thin last-scattering surface).  Thus for
secondary anisotropies, each $k$-mode can
contribute to a wide range of $l$'s.

\section{Secondary Anisotropies}
\label{sec:secondary}

After last scattering, the photons and baryons are no longer tightly coupled.
In fact, if the effects of reionization are negligible, there is no
coupling at all.  In this case, the photons simply propagate freely
along spacetime geodesics from last scattering to the observer.
The causes of secondary anisotropy are then entirely gravitational,
the dominant effect being the ISW effect.  Weak gravitational lensing
can also distort the anisotropy spectrum, although this effect is
generally small.

If the intergalactic medium reionized at a sufficiently early
redshift, then some fraction of the photons will interact again after
the time of ``last'' scattering.  The main result is that primary
fluctuations are erased, and in addition new fluctuations can be
generated from the new last-scattering surface.  However,
the last-scattering surface in a reionized model is extremely thick (since
the photon-baryon coupling is weak), so the nature of the regenerated
anisotropy is quite different from the primary anisotropy.

\subsection{Integrated Sachs-Wolfe Effect}

As Sachs \& Wolfe (1967) showed, fluctuations in
the spacetime curvature produce CMB anisotropy in two distinct ways.
The ``ordinary'' Sachs-Wolfe effect is simply the gravitational
red- or blueshift due to the potential difference between the points
of emission and reception of a photon.  In addition, if the gravitational
potential changes with time, there is an ``integrated'' Sachs-Wolfe
effect.  

Imagine a photon falling into a potential well, and
then climbing out the other side.  If the potential does not vary
with time, the photon suffers no net change in energy.  However,
if the potential well decays while the photon is passing through
it, then the redshift upon climbing out of the well is smaller
than the blueshift upon falling in.  The photon therefore gains
energy.  The magnitude of the ISW effect is given by an integral
along the photon's path:
\begin{equation}
\left(\Delta T\over T\right)_{ISW}=\int \left(\dot\Psi({\bf x},\eta)-
\dot\Phi({\bf x},\eta)\right)\,d\eta.
\end{equation}

We observed earlier that the gravitational potential is time-independent
if certain conditions are satisfied:
\begin{itemize}
\item[$\bullet$] The Universe is matter-dominated 
($\rho_{\rm matter}\gg\rho_{\rm rad}$).
\item[$\bullet$] Spatial curvature is negligible ($\Omega=1$).
\item[$\bullet$] Linear perturbation theory is valid ($\delta\ll 1$).
\end{itemize}
If all of these conditions are satisfied, there is no ISW effect.  However,
in any realistic cosmological model
some or all of these conditions are violated at some point.

\subsection{Early ISW Effect}

In a typical model, the epoch of matter-radiation equality occurs before
the time of last scattering, but not long before.  The matter-dominated
limit is therefore not quite correct around the time of last scattering
and shortly thereafter.  The decay in the potential shortly after
last scattering gives rise to the early ISW effect.
This effect is largest when the matter density $\Omega_0h^2$ is low.

The early ISW effect is most important 
on large scales.  Specifically, the scales
that are most affected are those with $k^{-1}$ comparable to
the time scale on which the potential decays.  Modes with wavelengths
much shorter than this oscillate many times while the potential
is decaying, causing both positive and negative ISW contributions,
which tend to cancel each other out.  The time scale for potential
decay is of order the horizon size at last scattering, so the early
ISW effect shows up on large angular scales $l\simless 200$.

\subsection{Late ISW Effect}

In models with $\Omega_0\ne 1$, the potential decays at late times,
typically at redshifts $z\simless\Omega_0^{-1}$.  This potential decay, which
occurs whether or not there is a cosmological constant, gives rise to
an ISW effect at late times.  As with the early ISW effect, modes
with wavelengths comparable to the time scale for the potential to
decay are most affected.  The relevant time scale is the horizon
size at the time of potential decay, so the late ISW effect
also leaves its imprint on large angular scales.

\subsection{Other ISW Effects}

At very late times, nonlinear structure forms, causing the potential
to grow with time.  The ISW effect due to nonlinear structure is
often called the {\em Rees-Sciama effect} (Rees \& Sciama 1968).  In
standard models, the Rees-Sciama effect is typically much weaker than
the other effects we have discussed (Seljak 1996a).

A background of primordial gravity waves, if there is one, produces
its own ISW effect.  Gravity waves redshift once they enter the
horizon, so modes that enter the horizon well before last scattering
leave no imprint on the CMB.  The gravity-wave contribution to the
CMB anisotropy therefore occurs on large angular scales $l\simless 100$.
Because of the quadrupolar nature of the spacetime distortion caused
by a gravity wave, the gravity-wave contribution to the CMB quadrupole
is enhanced relative to other modes.

There may be other sources of spacetime distortion besides
linear density fluctuations and gravity waves.  In particular,
topological defects cause spacetime curvature and hence an ISW
effect.  We will not discuss topological defects further; for
more information, see Paul Shellard's contribution to this volume,
and the references therein.

\subsection{Gravitational Lensing}

The ISW effect may be thought of as gravity imparting a ``kick'' to
a photon forward or backward along the direction of motion.  Gravity
can also kick the photons in the transverse directions, changing
their directions of motion but not their energies.  The result
of this weak gravitational lensing is that our image of the last-scattering
surface is slightly distorted, as if we were looking at it through
an irregular refracting medium.  This distortion of the last-scattering
surface results in a slight smearing of the angular power spectrum,
with power from the peaks being moved into the valleys.  The
effect is typically weak, resulting in changes at the few-percent
level in the power spectrum (Seljak 1996b).

\subsection{Reionization}

We will not undertake a detailed discussion of reionized models here.
Instead, we refer the interested reader to Roman Juszkiewicz's
contribution to this volume and references therein.  We will, however,
make some general comments.

The Gunn-Peterson test (Gunn \& Peterson 1965)
tells us that the intergalactic medium
is ionized out to redshifts of a few.  In CDM-like models of structure
formation, reionization is generally thought to occur at such
moderate redshifts, with the formation
of the earliest nonlinear structures.  If this is correct, then reionization
does not dramatically alter the CMB anisotropy predictions.
If, on the other hand, reionization somehow happened earlier, say
at $z\simgreat 100$, then a significant fraction of the CMB photons
have been scattered by the reionized matter after the so-called
epoch of last scattering.

\begin{figure}[t]
\centerline{\epsfig{file=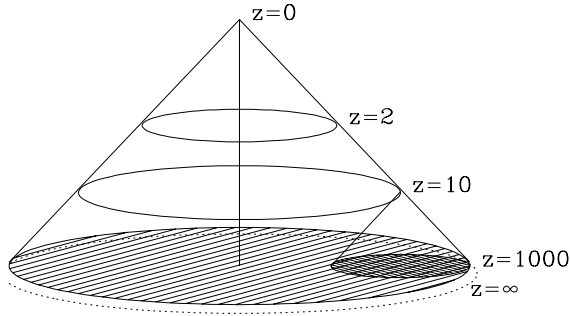,width=4in}}
\caption{
Our backward light cone.  The vertical axis represents conformal
time, and the horizontal axes are two of the three spatial
directions.  In the absence of reionization, each line of sight
corresponds to a particular point on the last-scattering surface
at $z\simeq 1000$.  In a reionized model in which a typical photon
last scattered at $z=10$, a photon arriving from a particular
direction may have originated from any point in the shaded circle.
Reprinted from Tegmark (1996c).
}
\label{fig:reion}
\end{figure}

The main effect of such early reionization is to erase anisotropy on
degree scales.  The reason is quite simple: if we have
early reionization, then a photon that comes toward us from a particular
direction need not have originated from that direction.  Rather, as
Figure \ref{fig:reion} illustrates, each direction on the sky contains
photons that originate from a variety of different locations
at the time of ``last'' scattering.  In severely reionized models,
the peaks are completely washed away.  Such models may already
be ruled out by degree-scale CMB experiments (Scott, Silk, \& White 1995).

Inhomogeneities and bulk motions of the reionized matter induce
new CMB anisotropies, which must generally be treated to second
order in perturbation theory (Ostriker \& Vishniac 1986; 
Hu, Scott, \& Silk 1994; Dodelson \& Jubas 1995), 
but we will not discuss these regenerated
anisotropies here.  We also neglect to discuss the effect of nonuniform
or patchy reionization, including the Sunyaev-Zel'dovich effect 
(Sunyaev \& Zel'dovich 1970).

\section{Summary of Anisotropy Formation}
\label{sec:summary1}

\begin{figure}[t]
\centerline{\epsfig{file=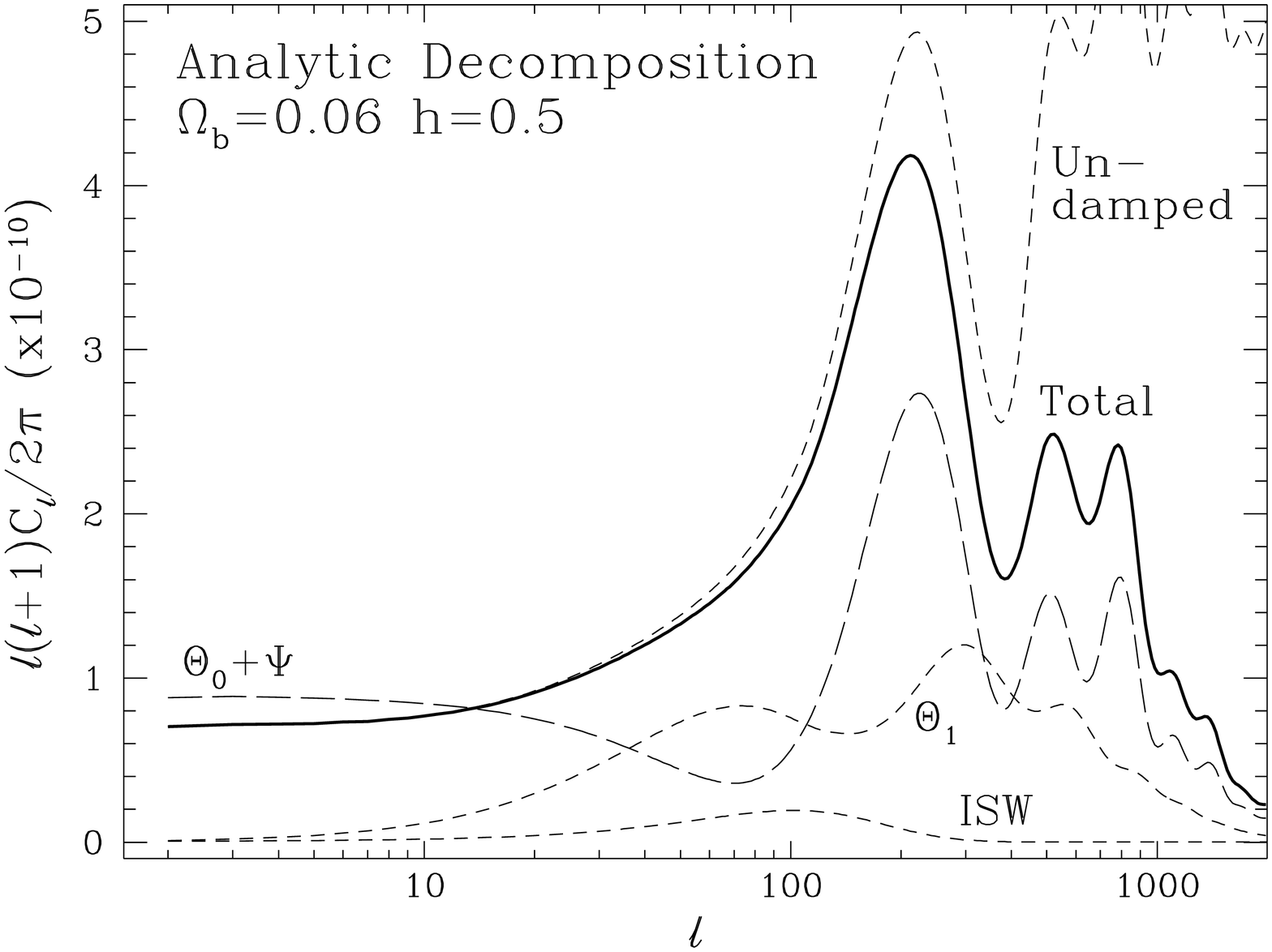,width=3in}}
\caption{
Analytic decomposition of anisotropies.  The solid line shows the
angular power spectrum of a critical-density CDM model.  The upper
dashed curve shows the spectrum that would be seen in the absence
of diffusion damping.  Note that the undamped peak heights increase
at scales small enough to have crossed the horizon before matter-radiation
equality.  (See Section \ref{sec:driving}.)
The other curves show the relative importance of the Sachs-Wolfe,
integrated Sachs-Wolfe, and Doppler ($\Theta_1$) contributions.
Reprinted from Hu (1996).
}
\label{fig:decomp}
\end{figure}

We have now concluded our tour of the mechanisms of anisotropy
formation.  Figure \ref{fig:decomp} illustrates some of the key
points.  The dominant features in a typical CDM power spectrum
are the peaks due to acoustic oscillations of the photon-baryon
fluid.  The peaks correspond to modes that are undergoing
maximum compression and
rarefaction at the time of last scattering.  Modes that are out of
phase with these modes produce anisotropy via the Doppler effect,
partially filling in the valleys between the acoustic peaks.
The effect of damping on small scales is evident, and the rise at $l\sim 500$
in the undamped spectrum shows the driving effect of the decaying
gravitational potential at early times.

We can use what we have learned to determine how the predicted anisotropy
spectrum should depend on the key cosmological parameters:

\begin{itemize}
\item[$\bullet$] In models with {\em spatial curvature} ($\Omega_0\ne 1$), 
the position of the acoustic peaks shifts due to geodesic deviation.
In addition, the late ISW effect boosts the large-scale power.
\item[$\bullet$] If there is a {\em cosmological constant}, then
the position of the peaks shifts slightly due to the increased distance
to the last-scattering surface, and again, the late ISW effect boosts
the large-scale power.
\item[$\bullet$] Lowering the {\em Hubble parameter} (for fixed $\Omega_0$)
reduces the matter density.  The gravitational driving of oscillations
is enhanced, and the peaks increase in height.
\item[$\bullet$] The higher the {\em baryon density} $\Omega_{\rm B}h^2$,
the greater the peak amplitude.  Odd-numbered peaks in particular are
enhanced.
\item[$\bullet$] The {\em spectral index} $n$ of the primordial
power spectrum essentially just tilts the angular power spectrum.
\item[$\bullet$] If we add {\em gravity waves} to a model, we
increase the quadrupole, and in addition the whole ``plateau''
at low $l$ rises relative to the acoustic peaks.
\end{itemize}

Although we have made many approximations in deriving these conclusions,
all of them are borne out by detailed Boltzmann calculations.  Because
the CMB anisotropy predictions depend sensitively on the various
parameters, an experiment that could map out the acoustic peaks
would be able to measure these cosmological
parameters accurately.  
The spatial curvature in particular should be relatively
easy to pick out, thanks to the shift in position of the first peak.
The relative positions of successive peaks also provide a robust
way of determining whether the initial conditions are isocurvature
or adiabatic.\footnote{For isocurvature initial conditions, the second
and third peaks occur at three and five times the location of the
first; in adiabatic models, they occur at two and three times.}
The dependence of the power spectrum on other parameters such as $h$
and $\Omega_{\rm B}$ is somewhat more subtle, but if we manage
to detect and measure the heights of two or three peaks, we should
be able to do quite well (Jungman \etal~1996), assuming, of course, that
the general paradigm sketched above is correct and the multiple peaks
are really there.

\section{Statistical Properties of $\Delta T/T$}
\label{sec:stats}

Before we discuss methods for comparing theories with data, we need to
discuss briefly the statistical properties of the CMB anisotropy
as it appears on the sky.  As we have mentioned, it is convenient
to expand the observed anisotropy in spherical harmonics:
\begin{equation}
{\Delta T\over T}(\hat{\bf r})=\sum_{l,m}a_{lm}Y_{lm}(\hat{\bf r}).
\label{eq:ylmexpand}
\end{equation}
We have focused on the anisotropy produced by an individual plane wave;
the observed anisotropy is of course a superposition of contributions
from all of these plane waves:
\begin{equation}
a_{lm}=\sum_{\bf k}a_{lm}^{({\bf k})}.
\end{equation}
Since all of the relevant physics is described by linear perturbation theory
--- as we know, everything in nature is linear\footnote{to first order.}
--- each $a_{lm}^{({\bf k})}$ is proportional to the initial density
perturbation $\delta_{\bf k}^{\rm (init)}$.  

One often assumes that the initial conditions
have ``random phases,'' meaning that different Fourier modes are
uncorrelated,
\begin{equation}
\langle\delta_{\bf k}\delta_{\bf k'}\rangle=0\qquad\qquad\hbox{when}
\qquad {\bf k}\ne {\bf k'}.
\end{equation}
In this case, the $a_{lm}^{({\bf k})}$ are also uncorrelated.
The mean-square power in a particular multipole is then simply the sum
of the contributions from the various Fourier modes:
\begin{equation}
\langle |a_{lm}|^2\rangle=\sum_{\bf k}\langle |a_{lm}^{({\bf k})}|^2\rangle.
\end{equation}
And, of course, the left-hand side of this equation is simply the
angular power spectrum $C_l$.  (This quantity is independent of
the azimuthal index $m$ as long as space is isotropic.)

We often go beyond the assumption of random phases and
assume {\em Gaussian initial conditions}.
This is a prediction of inflationary scenarios, but one often
assumes Gaussian initial conditions even in non-inflationary
phenomenological models such as isocurvature baryon models (Peebles~1987).
When we talk about a Gaussian theory, we simply
mean that at some initial time $t_i$ the density perturbation
$\delta$ was a realization of a Gaussian random field.  Bernard
Jones has provided a detailed discussion of Gaussian
random processes elsewhere in this volume; for our purposes, all
we need to know is that the assumption of Gaussian initial
conditions, together with homogeneity and isotropy, 
implies that each Fourier coefficient $\delta_{\bf k}$
is an independent Gaussian random variable of zero mean.\footnote{
That is, Gaussian initial conditions (together with homogeneity
and isotropy) imply random phases, but not
conversely.}
In other
words, the real-space density perturbation $\delta({\bf x})$ is
a stochastic superposition of plane waves of all different wavelengths.
Since a Gaussian random variable is completely determined by
its mean and variance, and since $\langle\delta_{\bf k}\rangle=0$,
the statistical properties of our Gaussian random field are completely
determined by
the power spectrum $P(k)\equiv\langle|\delta_{\bf k}|^2\rangle$.

If we assume Gaussian initial conditions, then each coefficient $a_{lm}$
is a Gaussian random variable, since it is a linear combination
of the Gaussian variables $\delta_{\bf k}$.  The statistical
properties of $\Delta T/T$ are therefore completely specified by
the means,
\begin{equation}
\langle a_{lm}\rangle=0,
\end{equation}
and the covariances,
\begin{equation}
\langle a_{lm}a_{l'm'}^*\rangle=C_l\delta_{ll'}\delta_{mm'},
\end{equation}
of the coefficients $a_{lm}$.  In other words, for Gaussian initial
conditions, the angular
power spectrum $C_l$ tells us everything we need to know.

Even when the initial conditions are not Gaussian, it often suffices
to treat the CMB anisotropy as Gaussian, at least on sufficiently
large angular scales.  The CMB fluctuation on large angular scales
is typically due to a superposition of many incoherent fluctuations.
Even if the individual fluctuations fail to be Gaussian, the central
limit theorem guarantees that the superposition will be approximately
Gaussian.  When comparing the COBE data with the predictions
of a cosmic string model, for example, it is perfectly adequate
to treat $\Delta T/T$ as Gaussian, even though the underlying perturbations
are highly non-Gaussian.

\section{An Introduction to CMB Data Analysis}
\label{sec:cmbdata}

\subsection{An Idealized Experiment}

We will explore the key issues in CMB data analysis by first considering
an absurdly idealized experiment (the sort of thing only a theorist
could dream up).  We will gradually introduce real-world complications
to see what the main issues are.

Imagine, then, an experiment that measured $\Delta T/T$ at many
pixels that cover the entire sky completely and uniformly.  Furthermore,
imagine that each data point is a perfect, noise-free measurement.
With this data set, we could determine each coefficient $a_{lm}$ with
essentially perfect accuracy by inverting equation (\ref{eq:ylmexpand}):
\begin{equation}
a_{lm}=\int {\Delta T\over T}(\hat{\bf r})Y_{lm}^*(\hat{\bf r})\,d\Omega
\approx {4\pi\over\npix}\sum_{p=1}^{\npix}{\Delta T\over T}(\hat{\bf r}_p)
Y_{lm}^*(\hat{\bf r}_p).
\label{eq:almestimate}
\end{equation}
Here $d\Omega$ is an element of solid angle in the direction of $\hat{\bf r}$,
$\npix$ is the total number of pixels, and $\hat{\bf r}_p$ is a unit
vector in the direction of the $p$th pixel.

Even in this hopelessly idealized experiment, we still can't measure
the angular power spectrum $C_l$ perfectly.  The reason is that $C_l$
is an ensemble-average quantity: it is the variance of the distribution
from which $a_{lm}$ is drawn.  We have only a finite number, $2l+1$,
samples of this distribution at each $l$.  This fact, generally
called {\em cosmic variance}, sets a fundamental limit on how
well we can ever hope to measure the angular power spectrum.\footnote{
Cosmic variance is closely related to the failure of ergodicity.
If $\Delta T/T$ were ergodic, then the average value of $|a_{lm}|^2$,
measured in different orientations over the sphere, would be the
ensemble-average quantity $C_l$.  But $\Delta T/T$ isn't ergodic,
so this doesn't work.}

If we assume Gaussian statistics, then the best estimator of $C_l$
is simply the average of $|a_{lm}|^2$ over $m$:
\begin{equation}
\hat C_l\equiv {1\over 2l+1}\sum_{m=-l}^l |a_{lm}|^2.
\end{equation}
This quantity is chi-squared distributed with $2l+1$ degrees of
freedom, and so it has a fractional uncertainty of
\begin{equation}
{\mbox{Var}^{1/2}(\hat C_l)\over C_l}=\sqrt{2\over 2l+1}.
\end{equation}
The unfortunate fact, therefore, is that even in a perfect experiment
we will never know $C_l$ with a fractional uncertainty better 
than $(l+{1\over2})^{-1/2}$.  We are stuck with a 63\% uncertainty
in the quadrupole power $C_2$ and a 30\% uncertainty in $C_{10}$,
although we can in principle hope to determine $C_{1000}$ to 0.3\%.

\subsection{Noise}
\label{sec:noise}

Let's mess up our nice, clean experiment by adding noise.  Each
pixel is no longer a perfect measurement of $\Delta T/T$: the
$i$th data point $d_i$ consists of a sum of signal and noise,
\begin{equation}
d_i={\Delta T\over T}(\hat{\bf r}_i)+n_i.
\label{eq:datadef1}
\end{equation}
Let us assume that the noise $n_i$ in each pixel is independent
and Gaussian distributed, with some standard deviation $\sigma$.
For the moment we will assume {\em homoskedasticity}, that is,
that $\sigma$ is the same in all pixels.

We can still try to estimate $a_{lm}$ using equation (\ref{eq:almestimate}),
\begin{equation}
\hat a_{lm}={4\pi\over\npix}\sum_{p=1}^\npix d_pY_{lm}^*(\hat{\bf r}_p),
\label{eq:almsum}
\end{equation}
and average over $m$ to get an estimate of $C_l$,
\begin{equation}
\hat C_l={1\over 2l+1}\sum_{m=-l}^l |\hat a_{lm}|^2,
\end{equation}
but this quantity will no longer be a good estimate of the true $C_l$;
it will be biased upward.  Using equations (\ref{eq:almsum}) and
(\ref{eq:datadef1}), together with the fact that $\langle n_pn_{p'}\rangle
=\sigma^2\delta_{pp'}$,
it is straightforward to check that
\begin{equation}
\langle |\hat a_{lm}|^2\rangle=C_l+{4\pi\over\npix}\sigma^2.
\end{equation}
The estimator $\hat C_l$ is the average of these quantities, so
it too is biased upward by $4\pi\sigma^2/\npix$.

We can of course get a better estimate of $C_l$ by subtracting off
the noise bias,
\begin{equation}
\hat C_l'\equiv\hat C_l-{4\pi\over\npix}\sigma^2.
\label{eq:hatcprime}
\end{equation}
We now have an unbiased estimator, but unfortunately the uncertainty
of $\hat C_l'$ has increased:
\begin{equation}
\mbox{Var}^{1/2}(\hat C_l')=\sqrt{2\over 2l+1}\left(
C_l+{4\pi\over\npix}\sigma^2\right).
\end{equation}

\subsection{A Digression on Statistical Methods in General}

The problem we just considered was a classic
example of statistical 
parameter estimation.  We had some {\em data}, $\{d\}$,
from which we wanted to estimate a {\em parameter}, $C_l$.  We did
it by choosing an {\em estimator}, $\hat C_l'$, which we could compute
from the data, and which we hoped would be close to the true value
of the parameter.

In the problem above, there was a fairly natural choice of an estimator,
but in general, for a more complicated problem, there may be no obvious
choice.  There is no universal, ``correct'' way to choose an estimator,
but in many situations the {\em maximum-likelihood} estimator is
a good choice.  We will illustrate maximum-likelihood estimators
with a simple example.

Suppose that we have $M$ data points $x_i$, each
of which is the sum of a signal
$s_i$ and some noise $n_i$.  We will take both $s_i$ and $n_i$
to be Gaussian random variables with zero mean.  The variances
of the signal and noise are
\begin{equation}
\langle s_i^2\rangle=S,\qquad\qquad\qquad
\langle n_i^2\rangle=N,
\end{equation}
And everything is uncorrelated:
\begin{equation}
\langle s_is_j\rangle=\langle n_in_j\rangle=\langle s_in_j\rangle=0,
\end{equation}
where the first two expressions assume $i\ne j$. Let us suppose
we know the noise variance $N$, and we want to estimate the
unknown quantity $S$, using a maximum-likelihood estimator.\footnote{
The astute reader will have noticed that this is precisely the
same problem we considered at the end of the last subsection.
We have simply changed all of the notation for no good reason.
To be specific, the correspondence with the
previous problem goes like this: $M\to 2l+1$, $S\to C_l$, $s_i\to a_{lm}$,
$N\to 4\pi\sigma^2/\npix$.
}

The first step is to compute the probability density of the
data for fixed $S$.  We want to know $p(\{x\}\ |\ S)$, where
$p(\{x\}\ |\ S)\,d^Mx$ is the probability of getting a set of data
that lie within an infinitesimal volume $d^Mx$ at the location
of the actual data $\{x\}$.  In this case, each $x_i$ is
an independent Gaussian with variance $S+N$,
\begin{equation}
p(x_i\ |\ S)={1\over\sqrt{2\pi(S+N)}}\exp\left(-x_i^2/2(S+N)\right),
\end{equation}
and the joint probability density is the product
\begin{eqnarray}
p(\{x\}\ |\ S)&=&\prod_{i=1}^Mp(x_i\ |\ s)\\
&=& \left(2\pi(S+N)\right)^{-M/2}\exp\left(-\sum_{i=1}^Mx_i^2\over
2(S+N)\right).
\end{eqnarray}

The probability density we have computed is a function of the
data $\{x\}$ for fixed $S$.  But the data are known, and $S$
is what we want to know.  We therefore choose to regard
this probability density as a function of $S$ and call it
the {\em likelihood}.
\begin{equation}
L(S)=p(\{x\}\ |\ S).
\end{equation}
When working with Gaussian probability distributions, it is often
convenient to work with the quantity ${\cal L}\equiv -2\ln L$
instead.  The maximum-likelihood estimator, as its name suggests,
is the value $\hat S$ of $S$ for which $L$ is maximized (or
$\cal L$ is minimized).  In other words, it is the value of the
parameter for which it would have been most likely for us
to get the data we actually did.

In the problem at hand, the maximum-likelihood estimator is found
by differentiating
\begin{equation}
{\cal L}=M\ln2\pi+M\ln(S+N)+{\sum x_i^2\over S+N}
\end{equation}
with respect to $S$,
setting the result equal to zero, and solving for $S$.  The
result is
\begin{equation}
\hat S={1\over M}\sum_{i=1}^M x_i^2-N.
\end{equation}
That is, we compute the mean-square value of the data points and
subtract of the noise bias.  This is precisely what we did when
we computed $\hat C_l'$ in equation (\ref{eq:hatcprime}).
Although we didn't know it at the time, 
we were using a maximum-likelihood estimator.

In this case, the maximum-likelihood estimator turned out to be
unbiased: its ensemble average $\langle \hat S\rangle$ is
equal to the correct value $S$.  In general, there is no guarantee
that this will happen.  To take a simple example, suppose that
we had chosen to estimate the quantity $S^{289}$ instead of $S$.
The maximum-likelihood estimator would be $\hat S^{289}$, and it
is easy to see that this quantity is highly positively biased.

Now we know how to estimate parameters.  But in most cases an
estimator isn't much good without a way of quantifying the uncertainty
in it.  Methods for doing this generally fall into two categories:
the classical or {\em frequentist} approach ({\it e.g.},
Rice 1995) and the {\em Bayesian}
approach ({\it e.g.}, Berger 1985, Gull \& Daniell 1978, Press 1996).  
We will discuss each in turn.

In the frequentist picture, we look at one value of the
parameter $S$ at a time, and try to determine if that value
is so far from our estimator $\hat S$ that it is ruled out.
Specifically, 
for each $S$, we compute the probability distribution of the
estimator $\hat S$.  We use this probability distribution
to determine how likely it is that we would have gotten a value
of $\hat S$ as far off as we did, or worse.  If the actual
value of $\hat S$ is far off in the tail of the probability distribution,
then this probability will be low.  If the probability lies
below some {\em significance level} (say 5\%), we say that
that value of $S$ is ruled out with 5\% significance.\footnote{
Astrophysicists
often phrase that same statement differently, saying that the
value is ruled out at {\em 95\% confidence}.}
We repeat this process for a range of values of $S$, and we 
say that the set of values that are not ruled out form
a 95\% {\em confidence interval} for the parameter.

For a frequentist, a value of the parameter is ruled out if
there is a low probability of getting data that fits as badly as the
actual data.  The Bayesian approach is quite different in spirit:
A Bayesian attempts to determine the subjective probability
distribution that characterizes her knowledge of the parameter
given the data.  Armed with that probability distribution, 
she can calculate how likely the parameter is to lie in any
particular range.

In order to implement the Bayesian strategy, we want to turn
the likelihood function $L(S)=p(\{x\}\ |\ S)$, which represents the probability
of the data given a value of the parameter, into $p(S\ |\ \{x\})$, the
probability of the parameter given the data.  The way to do this
is to invoke Bayes's theorem:
\begin{equation}
p(S\ |\ \{x\})\propto p(\{x\}\ |\ S)p(S),
\label{eq:bayes}
\end{equation}
with the constant of proportionality chosen to make the integral
of the left-hand side equal one.
The left-hand side of this equation is the {\em posterior
probability distribution}, and it is precisely what we are looking
for: it tells us the probability of a particular parameter value,
given the data.  On the right-hand side we have the product of the
likelihood function and the {\em prior distribution} of the parameter
$S$.  The latter represents our state of knowledge of
$S$ before we looked at the data.

A Bayesian characterizes the uncertainty in a parameter estimate
by drawing a {\em credible region} around the estimate.  A 95\%
credible region, for example, is an interval $S_{\rm min}<S<S_{\rm max}$ 
such that there is a 95\% posterior probability that $S$ lies in that
interval,
\begin{equation}
\int_{S_{\rm min}}^{S_{\rm max}} p(S\ |\ \{x\})\,dS=0.95.
\end{equation}
The boundaries $S_{\rm min}$ and $S_{\rm max}$ of the credible
region are typically chosen to have equal values of the posterior
probability density.

Although the frequentist approach is the one most people think
of when they think of statistics, and although most scientists
profess to prefer it, many if not most error bars in cosmology
are determined using Bayesian techniques.

The main objection people raise to the Bayesian is that the
final results depend on the prior distribution $p(S)$.  For a true,
orthodox Bayesian, this is not really a problem: the Bayesian
view is that all probabilities represent our subjective knowledge,
and that prior distributions are therefore secretly built into
all statistical reasoning.  It is better, the argument goes,
to have the prior out in the open for all to see.

Whether or not you like this argument, there is no denying that 
in practice choosing a prior can be tricky.  If one has essentially
no prior knowledge about the parameter, then the prior distribution
should be broad and flat.  [For a flat prior, we
can see from equation (\ref{eq:bayes}) that the posterior
probability distribution is simply the likelihood function.]
But even in this situation, it is not generally obvious which ``flat''
prior to choose.  For example, if we are trying to estimate an element
of the power spectrum $C_l$, should we choose a prior that is flat
in $C_l$ or one that is flat in $\sqrt{C_l}$?  ($C_l$ is after all
a mean-square amplitude; maybe the r.m.s. amplitude is a more ``natural''
choice.)  Perhaps we should even choose a prior that
is flat in $\ln C_l$, since such a prior avoids choosing
a preferred scale.  It would be hard to say that
any of these choices is ``wrong,'' but in some situations
the result of a calculation may depend on which choice
is made.  For
an example, see Bunn \etal~(1994).  

The situation is not as bad as it appears, however.  If the 
data set in question contain a good, strong detection
of the parameter of interest, then the likelihood function
is sharply peaked, and the shape
of the posterior probability (\ref{eq:bayes}) is determined
mostly by the likelihood rather than the prior.  Prior
dependence is thus typically weak in the case of strong
detections.  The situations where prior dependence is a serious
problem are typically those in which someone is trying
to coax a value out of a data set that is capable of
only a weak constraint anyway.

\subsection{Incomplete Sky Coverage}

We now return to our hypothetical CMB experiment.  The next
complication we need to consider has to do with the fact that no
actual experiment ever achieves complete sky coverage.  In the case of
COBE, pixels close to the Galactic plane are contaminated, leaving only
about two thirds of the sky usable.  All other experiments to date
have covered even smaller patches of sky.

This fact requires us to completely change our approach.  As much as
we would like to estimate each $a_{lm}$ and hence each $C_l$
individually, in the absence of complete sky coverage it is impossible
to do so.  There is in fact no estimator of a
particular $C_l$ that is ``uncontaminated,''
{\it i.e.}, that is independent of all of the other $C_{l'}$.

We may decide that it is important to estimate each $C_l$
individually, with the minimum possible contamination from other
multipoles.  Tegmark (1996a) has devised power-spectrum estimators 
with this property in mind and has applied them to both galaxy
surveys (Tegmark 1995) and the four-year COBE data (1996b).
For instance, suppose we have our
hearts set on knowing the value of $C_{17}$ as well as possible.
Since the power spectrum is quadratic in $\Delta T/T$, it is
natural to choose a quadratic estimator,
\begin{equation}
\hat C_{17}=\sum_{i,j}A_{ij}d_id_j-B.
\end{equation}
Here $d_i$ is a data point and we want to choose the matrix 
elements $A_{ij}$ and the bias correction $B$ in order to get
as good an estimator as possible.  Tegmark (1996a) proposes that we
choose these quantities to make our estimator unbiased and to
minimize the dependence of $\hat C_{17}$ on all of the other
$C_l$'s.  He shows that it is impossible to completely remove
contamination from other multipoles and that in general 
the ``spectral resolution'' $\Delta l$ of an experiment
is approximately the reciprocal of the angular scale $\Delta\theta$
covered by the sky map.
In particular, for an experiment like COBE, $\Delta\theta\sim 1$
radian, and it turns out that it is
possible to estimate a particular $C_l$ with significant contamination 
only from modes with $\Delta l\approx 2$ (Tegmark 1996a, 1996c).

\subsection{Maximum-Likelihood Parameter Estimation}

We may, however, decide that it isn't so important to estimate
each $C_l$ individually.  Often, a more fruitful approach is
to parameterize the power spectrum $C_l$ with a small number $k$
of parameters,
\begin{equation}
C_l=C_l(q_1,q_2,\ldots,q_k),
\end{equation} 
and use maximum-likelihood methods to estimate
those parameters.  This is in fact the usual approach in CMB
data analysis.  Specific choices of the parameters $\{q\}$
include the following:
\begin{itemize}

\item[$\bullet$] We may assume a {\em shape} for the power spectrum
and estimate the {\em normalization}.  In this case, there is only
one free parameter, which is conventionally taken to be the
quadrupole amplitude $\langle Q\rangle\equiv\sqrt{5C_2/4\pi}$.\footnote{
A bewildering variety of notations exist in the literature.  We choose
to call this quantity $\langle Q\rangle$ to emphasize the fact
that it is a theoretical ensemble-average quantity.  In particular,
it is not
the same as the local quadrupole $Q_{\rm rms}\equiv
\sum_{m=-l}^l |a_{2m}|^2/4\pi$.  The COBE group generally denotes its
estimators of $\langle Q\rangle$ by $Q_{\rm rms-PS}$.}
Most degree-scale experiments are only powerful enough to determine
a single number, the total power.  One therefore frequently assumes
a ``flat'' power spectrum $l(l+1)C_l=\mbox{const.}$ and estimates
the normalization, which in this context is often called $Q_{\rm flat}$.

\item[$\bullet$] Both the normalization $\langle Q\rangle$ and the
spectral index $n$ may be chosen as free parameters.  For a large-angle
experiment like COBE, the predicted power spectrum depends only weakly
on many of the other parameters.

\item[$\bullet$] White \& Bunn (1995) have suggested a phenomenological
parameterization of the power spectrum.  At large angular
scales, many popular
theoretical models are well approximated by power spectra that are
quadratics in $\log l$.  To be specific, we may set
\begin{equation}
l(l+1)C_l=D_1(1+D'(\log_{10}l-1)+\hbox{$1\over 2$}D''(\log_{10}l-1)^2)
\end{equation}
and work with a three-parameter family $(D_1,D',D'')$ of power spectra.

\item[$\bullet$] We may choose to divide the power spectrum
over the range probed by a particular experiment into a small number
of ``bands.''  We then estimate the power in each band,
assuming that $l(l+1)C_l$ is constant
in each band.  This has been done for COBE (Hinshaw \etal~1996) 
and Saskatoon (Netterfield \etal~1996), although the latter uses
a completely different method.

\end{itemize}

No matter what parameterization we adopt, we need a way to
compute the likelihood $L$ for a given power spectrum.  As long
as we assume Gaussian statistics, it is relatively easy to write down
a formula for the likelihood, although as we shall see it can be
cumbersome to compute it in practice.

We begin by introducing some notation.  Each data point $d_i$ is
as usual the sum of the signal $\Delta T/T(\hat{\bf r}_i)$ and
noise $n_i$.  Expanding $\Delta T/T$ in spherical harmonics, we have
\begin{equation}
d_i=\sum_{l,m}a_{lm}Y_{lm}(\hat{\bf r}_i)+n_i.
\label{eq:datadef}
\end{equation}
Let us denote a pair of indices $(lm)$ by a single Greek index $\mu$.
The correspondence is $\mu=l(l+1)+m$, so that $\mu$ ranges from 1
to $\infty$ as $(lm)$ take on all of their allowed values.  Then
we can write equation (\ref{eq:datadef}) more compactly as
\begin{equation}
\vec d={\bf Y}\vec a+\vec n,
\end{equation}
where $\vec d=(d_1,d_2,\ldots,d_\npix)$ is the data vector,\footnote{
We denote vectors that live in abstract spaces such as ``pixel
space'' by arrows, and vectors in real three-dimensional space 
are written in boldface.}
$\vec n=
(n_1,\ldots,n_\npix)$ is the noise vector, and the infinite-dimensional
vector $\vec a=(a_1,a_2,\ldots,a_\mu,\ldots)$ contains the spherical
harmonic coefficients.  The $\npix\times\infty$-dimensional
spherical harmonic matrix $\bf Y$ has
elements
\begin{equation}
Y_{i\mu}=Y_\mu(\hat{\bf r}_i).
\end{equation}

The statistical properties of $\vec d$ are determined by the properties
of $\vec a$ and $\vec n$.  Assuming Gaussian statistics, both are
Gaussian random vectors with zero mean and covariances given by
\begin{eqnarray}
\langle a_\mu a_\nu^*\rangle&=&C_\mu\delta_{\mu\nu}\equiv C_{\mu\nu},\\
\langle n_in_j\rangle&=&\sigma_j^2\delta_{ij}\equiv N_{ij},\\
\langle a_\mu n_i\rangle&=&0.
\end{eqnarray}
($C_\mu\equiv C_l$ where $l$ is the index corresponding to $\mu$, and
$\bf C$ and $\bf N$ are diagonal matrices.)
Since $\vec d$ is a linear combination of $\vec a$ and $\vec n$, it
too is a multivariate Gaussian, and the likelihood function therefore
has the form
\begin{equation}
L(C_l)\equiv p(\vec d \ |\  C_l)=
{1\over (2\pi)^{\npix/2} \det^{1/2}{\bf M}} 
\exp\left(-\hbox{$1\over 2$}\,\vec d\transpose
{\bf M}^{-1}\vec d\,
\right).
\label{eq:gausslikely}
\end{equation}
The T denotes a transpose, and the covariance matrix $\bf M$ is given
by
\begin{equation}
{\bf M}\equiv\langle \vec d \vec d\transpose\rangle
=\langle({\bf Y}\vec a+\vec n)({\bf Y}\vec a+\vec n)^{\rm T}\rangle
={\bf YCY}^{\rm T}+{\bf N}.
\label{eq:covarmat}
\end{equation}

In principle, we are now ready to estimate parameters.  Equation
(\ref{eq:gausslikely}) tells us how to compute the likelihood
for any particular power spectrum $C_l$, so all we need to do
is hunt through our parameter space for the parameters that
maximize the likelihood.

In fact, for a typical degree-scale experiment with tens or at
most hundreds of pixels, this is essentially what is done.  For
a large data set such as COBE, though, there are too many pixels for
this to be convenient: each time we wish to compute a likelihood,
we must invert the $\npix\times\npix$ matrix $\bf M$.  For COBE,
therefore, we must implement some form of ``data compression''
to make the analysis tractable.  (Data compression
will be even more essential for a future satellite experiment
with orders of magnitude more pixels than COBE.)

\subsection{Beam-Smoothing and Chopping}

Before we discuss data compression, though, we need to discuss one
more issue.  The hypothetical experiment we have been discussing
is still overly idealized in one important way.  We have assumed
that the signal measured by the experiment is the temperature
anisotropy $\Delta T/T$ at a point.  In reality, no experiment
has perfect resolution, so the observed signal is actually the
convolution of $\Delta T/T$ with some beam pattern or point-spread
function.  Furthermore, many experiments chop their beams between
two (or more) points on the sky, with the measured signal
being a difference between these points.

The effect of the beam pattern on our analysis is fairly simple.
Let $B(\alpha)$ represent the response of the instrument to
a point an angular distance $\alpha$ from the line of sight.
(We assume that the
beam pattern is azimuthally symmetric.)  Then what the experiment
actually measures is the convolution of the anisotropy with
the beam pattern,
\begin{equation}
\left({\Delta T\over T}\star B\right)=\sum_{l,m}\bar a_{lm}Y_{lm}.
\end{equation}
The coefficients $\bar a_{lm}$ are related to the true anisotropy
coefficients $a_{lm}$ like this:\footnote{
This result is simply the spherical version of the convolution
theorem for Fourier transforms, $\widetilde{f\star g}=\tilde f\tilde g$.}
\begin{equation}
\bar a_{lm}=B_la_{lm},
\end{equation}
where $B_l$ is the expansion in Legendre polynomials of $B$,
\begin{equation}
B_l=\int_{-1}^1d\cos\alpha\,B(\alpha)P_l(\cos\alpha).
\end{equation}
If the beam pattern happens to be a Gaussian,
\begin{equation}
B(\alpha)\propto \exp(-\alpha^2/2\sigma^2),
\end{equation}
then the Legendre coefficients are
\begin{equation}
B_l=\exp(-\hbox{$1\over2$}\sigma^2 l(l+1)).
\end{equation}
Note that as expected
$B_l$ is very small for $l\gg \sigma^{-1}$, {\it i.e.},
for angular scales $\theta \ll \sigma$.

We can adapt all of the previous results of this section to take
beam-smoothing into account by simply saying that our experiment
is measuring the beam-smoothed power spectrum,
\begin{equation}
\bar C_l\equiv C_lB_l^2,
\end{equation}
instead of $C_l$.

We can account for the effect of beam-switching in a similar way.
Consider an experiment that chops between two points with spherical
coordinates $(\theta,\phi+{1\over2}\alpha)$ and $(\theta,\phi-{1\over 2}
\alpha)$.  Ignoring beam-smoothing, 
the observed signal $d$ is the difference in the 
anisotropy between these two points:
\begin{eqnarray}
d&=&{\Delta T\over T}(\theta,\phi+\hbox{$1\over 2$}\alpha)-
{\Delta T\over T}(\theta,\phi-\hbox{$1\over 2$}\alpha)\\
&=&\sum_{l,m}a_{lm}\left(Y_{lm}(\theta,\phi+\hbox{$1\over 2$}\alpha)
-Y_{lm}(\theta,\phi-\hbox{$1\over 2$}\alpha)\right).
\end{eqnarray}
The azimuthal dependence of $Y_{lm}$ is $\exp(im\phi)$, so
\begin{eqnarray}
d&=&\sum_{l,m}a_{lm}Y_{lm}(\theta,\phi)\left(\exp(\hbox{$1\over 2$}im\alpha)-
\exp(-\hbox{$1\over 2$}im\alpha)\right)\\
&=&\sum_{l,m}a_{lm}Y_{lm}(\theta,\phi)2i\sin\hbox{$1\over 2$}m\alpha.
\end{eqnarray}
The net result is that $a_{lm}$ is replaced by $2ia_{lm}\sin{1\over2}m\alpha$,
so modes with low $|m|$ are suppressed.  Since $m$ ranges from $-l$ to $l$,
this suppression affects primarily modes with low $l$.\footnote{
The fact that modes with low $|m|$ are suppressed depends on the
fact that we have oriented our coordinate system with the chop
in the azimuthal direction.  In contrast, the statement that, on average,
modes with low $l$ are suppressed is independent of the orientation
of the coordinate system.}

This suppression is conventionally quantified by computing a ``window
function'' that represents the sensitivity of the experiment to
different multipoles.  To do this, we compute the mean-square signal,
\begin{equation}
\langle d^2\rangle=\sum_{l,m}C_l |Y_{lm}(\theta,\phi)|^2 \left(2\sin
\hbox{$1\over 2$}m\alpha\right)^2\equiv
\sum_l \left(2l+1\over 4\pi\right)C_l W_l.
\label{eq:chop}
\end{equation}
The window function $W_l$ is small for low $l$, indicating that
chopping has rendered this experiment insensitive to the largest
angular scales.

Note that we have not included beam-smoothing in equation (\ref{eq:chop}).
The correct window function, including beam-smoothing, is obtained
by multiplying this result by $B_l^2$.

Equation (\ref{eq:chop}) gives the window function for the particularly
simple case of a single-difference experiment.  There are more complicated
switching strategies, including sinusoidal chops and triple-beam experiments.
For a more detailed discussion of window functions, see White \& Srednicki
(1995).

\section{Likelihood Analysis of the COBE Data}
\label{sec:cobe}

In the previous section, we discussed various issues of CMB data analysis
from a general point of view.  We will now apply what we have learned
to a specific example, namely the COBE DMR data.  We will not
describe the COBE instrument in detail; the interested reader is
referred to George Smoot's contribution to this volume, as well
as to the papers reporting the four-year DMR data (Bennett \etal~1996,
G\'orski \etal~1996, Hinshaw \etal~1996, Banday \etal~1996)
and references therein.  We will content ourselves with
mentioning a few of the most relevant facts.

The COBE DMR produced all-sky maps of the microwave radiation
at three frequencies, 31 GHz, 53 GHz, and 90 GHz, with
a beam size of $7^\circ$ (FWHM).  The maps consist of 6144 pixels,
although only about 4000 of them are at high enough Galactic
latitude to be used for studying the CMB.  Although
the DMR is a differencing instrument, the data
have been used to produce sky maps of $\Delta T/T$, so we do not
need to worry about beam-switching in our analysis.  We do, however,
have to worry about the fact that the maps are insensitive
to the monopole and dipole of the anisotropy.\footnote{
Actually, COBE is in principle perfectly sensitive to the dipole;
however, the intrinsic CMB dipole is impossible to distinguish from
the much larger dipole due to our own motion with respect to
the CMB center-of-momentum frame.}

\begin{figure}[t]
\centerline{\epsfig{file=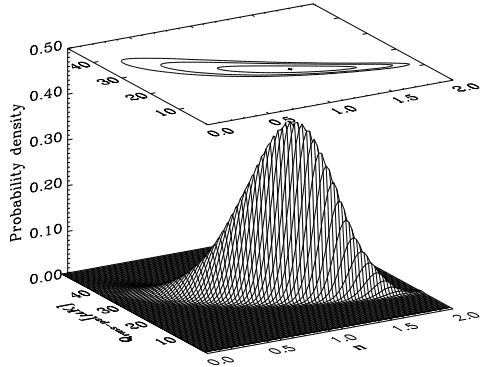,width=3in}}
\caption{
The likelihood function for the two-year COBE DMR data, based on a
brute-force analysis involving the entire pixel covariance matrix.
Only the Sachs-Wolfe contribution to the anisotropy is included.
See Tegmark \& Bunn (1995) for further details.
}
\label{fig:brute}
\end{figure}

The noise in the COBE maps appears to be Gaussian, and different pixels
have noise that is approximately uncorrelated (Lineweaver \etal~1994).
Therefore, as long as the CMB anisotropy obeys Gaussian statistics,
equation (\ref{eq:gausslikely}) applies:
\begin{equation}
{\cal L}\equiv -2\ln L=\ln\left((2\pi)^\npix\det{\bf M}\right)+
\vec d\transpose{\bf M}^{-1}\vec d,
\end{equation}
where ${\bf M}={\bf Y}\bar{\bf C}{\bf Y}^{\rm T}+{\bf N}$
with $N_{ij}=\sigma_i^2\delta_{ij}$.
The matrix $\bf M$ is $\sim4000\times 4000$, which is a size that
can be inverted, with sufficient patience, on a workstation.
Tegmark \& Bunn (1995) have performed such a brute-force analysis
on the two-year COBE DMR data for a two-parameter family of
power spectra, with results shown in Figure \ref{fig:brute}.
However, if we wish to explore a larger parameter space, we
must find a more efficient way to compute likelihoods.

\subsection{Data Compression}

All likelihood analyses of the COBE data, with the exception
of the brute-force analysis mentioned above, have involved some
form of data compression.  That is, the pixel data $\vec d$
has been mapped to some smaller-dimensional data vector,
which has been used for computing likelihoods.  We will focus
on {\em linear} methods of data compression, in which
the compressed data vector $\vec x$ is linear in $\vec d$,
\begin{equation}
\vec x={\bf A}\vec d,
\end{equation}
for some $K\times\npix$ matrix $\bf A$ with $K<\npix$.  $\vec x$
is a Gaussian random vector, so we can use equation (\ref{eq:gausslikely})
to compute the likelihood
of $\vec x$ in terms of the covariance matrix
\begin{equation}
{\cal M}\equiv\langle\vec x\vec x^{\rm T}\rangle={\bf AMA}^{\rm T}.
\end{equation}
Of course, the likelihood computed in this way will not be the same
as the true likelihood computed from $\vec d$, but we can hope
that, if we perform our data compression wisely, we will get a reasonable
approximation to the true likelihood.

In effect, linear data compression is equivalent to expanding the
sky map in a set of normal modes, namely the rows of $\bf A$.
Each element of the compressed data vector $\bf x$ is approximately
the integral of the sky map, multiplied by some function: heuristically,
we can write
\begin{equation}
x_i=\sum_{j=1}^\npix A_{ij}d_j\approx \int A_i(\hat{\bf r})d(\hat{\bf r})
\,d\Omega.
\end{equation}
If our pixels uniformly covered the whole sky, we would choose these
mode functions to be the spherical harmonics by setting
$A_{\mu j}=Y_\mu(\hat{\bf r}_j)$.  Then $x_\mu$ would be an estimate
of $a_\mu$ (up to an overall normalization).  
In fact, we would be performing precisely the
analysis described in Section \ref{sec:noise}.

Even though we do not actually have complete sky coverage, there
is still nothing stopping us from choosing the rows of $\bf A$
to be the spherical harmonics.  This is in fact the technique
described by G\'orski (1994), which has been applied to the DMR
data by G\'orski \etal~(1994, 1996).\footnote{
G\'orski's method involves the additional step of orthogonalizing
the spherical harmonics with an algorithm like Gram-Schmidt.  Orthogonalizing
with respect to the monopole and dipole is an excellent way to
render the data insensitive to these modes, but orthogonalizing
the modes with $l\ge 2$ with respect to each other has no effect
on the likelihoods.}
By cutting off the spherical harmonic expansion at $l=30$, G\'orski
\etal\ compress the data from $\sim4000$ to $\sim1000$ numbers,
with little loss of cosmological information.  This is possible
because the cosmic signal in the data drops off rapidly with
increasing $l$ (due to both the beam cutoff and the shape of
the anisotropy power spectrum), while the noise has approximately
equal power in all modes.

\subsection{The Karhunen-Lo\`eve Transform}

The Karhunen-Lo\`eve transform (Karhunen~1947), which is also
known as optimal subspace filtering or expansion in signal-to-noise
eigenmodes, is another prescription for linear data compression.
It was first introduced to CMB data analysis by Bond (1994, 1995, 1996), 
and
has been used extensively on the COBE data (Bunn, Scott, \& White~1995;
Bunn 1995; White \& Bunn 1995; 
Bunn \& Sugiyama 1996; 
Bunn, Liddle, \& White 1996; Bunn \& White 1996) 
as well as in analyzing galaxy
catalogues (Vogeley \& Szalay 1996).

Let us consider a one-parameter family of power spectra $C_l(q)$,
where the true value of $q$ is $q_0$.  We wish to choose our
method of data compression ({\it i.e.}, the matrix $\bf A$)
to enable us to estimate $q$ as well as possible.  Specifically,
we choose $\bf A$ to maximize our ability to reject incorrect
values of $q$.

On average, the likelihood function $L(q)$ has a peak at the true
value $q=q_0$, so $\langle L'(q_0)\rangle=0$.  The average
rejection power is determined by the rate at which the likelihood
declines when we move away from this peak.  The figure of merit
for describing rejection power is therefore
\begin{equation}
\gamma\equiv\left<\left.{d^2{\cal L}\over dq^2}\right|_{q=q_0}\right>.
\end{equation}
The Karhunen-Lo\`eve transform consists of choosing the compression
matrix $\bf A$ to maximize $\gamma$ (for a fixed value of $K$, the
dimension of the compressed data vector).

To solve this optimization problem, we write down the likelihood
in terms of the reduced data vector $\vec x$,
\begin{equation}
{\cal L}=K \ln 2\pi +\mbox{Tr}\left(\ln({\bf AMA}^{\rm T})+
({\bf AMA}^{\rm T})^{-1}\vec x\vec x^{\rm T}\right).
\end{equation}
Then we compute $\gamma$, vary a matrix element $A_{ij}$, and set
$\delta\gamma=0$.  After some algebra, we find that each row $\vec\alpha_a$
of $\bf A$ must satisfy an eigenvalue equation,
\begin{equation}
{\bf M}'_0\vec\alpha_a=\lambda_a{\bf M}_0\vec\alpha_a.
\label{eq:eigen}
\end{equation}
Here ${\bf M}_0$ is the covariance matrix $\bf M$ corresponding
to the correct parameter value $q=q_0$, and
\begin{equation}
{\bf M}'_0=\left.{d{\bf M}\over dq}\right|_{q=q_0}.
\end{equation}
The rejection power $\gamma$ is simply the sum of the squares of the
eigenvalues $\lambda_a$.

This completes our prescription for choosing the matrix $\bf A$.  We should
choose the rows of $\bf A$ to be the solutions of equation (\ref{eq:eigen})
with the largest values of $|\lambda_a|$.  Furthermore, we know when
it is safe to stop adding new rows: once all of the
remaining eigenvalues $\lambda_a$ are small, we will no longer significantly
increase $\gamma$ by adding more rows to $\bf A$.

To get an intuitive understanding of the Karhunen-Lo\`eve transform,
consider the case where the parameter $q$ is the normalization of
the power spectrum, so $C_l(q)=qC_l^{(0)}$.  Then we can rewrite
the eigenvalue equation (\ref{eq:eigen}) as
\begin{equation}
{\bf M}_{\rm signal}\vec\alpha_a=\hat\lambda_a{\bf M}_{\rm noise}\vec\alpha_a,
\label{eq:eigen2}
\end{equation}
where $\hat\lambda_a=\lambda_a/(1-\lambda_a)$, and 
${\bf M}_{\rm signal}={\bf Y}\bar{\bf C}{\bf Y}^{\rm T}$ and 
${\bf M}_{\rm noise}=\bf N$ are the signal and noise contributions to $\bf M$.
We can see from equation (\ref{eq:eigen2}) that $\vec\alpha_a$ is
an eigenvector of ${\bf M}_{\rm noise}^{-1}{\bf M}_{\rm signal}$.
This is why Bond (1994, 1995) calls it an ``eigenmode of the signal-to-noise
ratio.''  In effect, the Karhunen-Lo\`eve transform tells us
which directions in the $\npix$-dimensional pixel space are most
sensitive to the cosmic signal, and which are dominated by noise.

The reader, being an extraordinarily perceptive soul, is no doubt
wondering at this point whether this whole procedure is worth
the trouble.  After all, our original goal was to avoid having
to invert an $\npix\times\npix$ matrix.  Now we find ourselves having
to solve an $\npix$-dimensional eigenvalue problem, which is much harder
than simply inverting a matrix.  Recall, however, that our objection
to a brute-force likelihood analysis was that we didn't want to
invert the large matrix $\bf M$ {\em repeatedly} as we varied the
power spectrum.  The Karhunen-Lo\`eve eigenvalue problem needs to
be solved only once, with all future operations being performed on
the $K$-dimensional compressed data vector.  Furthermore, it
turns out that we can save ourselves a lot of work by solving
equation (\ref{eq:eigen}) in spherical harmonic space rather than
real space (Bunn 1995).  Once we choose some cutoff $l_{\rm max}$, the
dimension of the eigenvalue problem is reduced from $\npix$ to $\sim
l_{\rm max}^2$.  It turns out that none of the high signal-to-noise
eigenmodes have significant power beyond $l=30$ or so, so we can
safely choose $l_{\rm max}$ to be 40 or 50, resulting in a substantial
saving in computational effort.

The Karhunen-Lo\`eve transform depends on a choice
of power spectrum.  Ideally, we would like to use the true power
spectrum, but of course we don't know the true power spectrum.\footnote{
If we did, there would be no need to perform the analysis!}
We must therefore choose a {\em fiducial power spectrum} more or
less arbitrarily.  In principle, this could lead to trouble: we
might find that the choice of fiducial power spectrum had a significant
effect on our final results.  There are two ways to address this question:
we can repeat the analysis with different fiducial power spectra,
and we can perform Monte Carlo simulations to check that the
likelihood analysis returns unbiased estimates of the parameters of
interest.\footnote{
Even in methods that do not involve a choice of fiducial power spectrum,
it is wise to perform simulations to test for bias.  Even a
brute-force likelihood analysis using the full pixel data is not
guaranteed to return unbiased parameter estimates.}

In the case of the COBE data, extensive tests have
revealed that sensitivity to the fiducial
power spectrum is not a problem (Bunn 1995, Bunn \& White 1996).
For example, the maximum-likelihood normalization of an $n=1$
Sachs-Wolfe spectrum is $\langle Q\rangle=18.73\pm 1.25\,\mu\rm K$
using an $n=1$ fiducial power spectrum and $\langle Q\rangle=18.74\pm
1.25\,\mu\rm K$ using an $n=1.5$ power spectrum.  The maximum-likelihood
value of $n$ also does not change when we change the fiducial power
spectrum.  Furthermore, Monte Carlo simulations show that our
estimates of $\langle Q\rangle$ and $n$ are unbiased to an accuracy
much better than the statistical uncertainty 
($\simless 0.03\,\mu\rm K$ and $\simless 0.05$ 
respectively).  See Bunn \& White
(1996) for further details.

\subsection{Monopole and Dipole Removal}

Since the COBE data do not contain useful monopole and dipole
information, it is customary to remove a best-fit monopole and dipole
from the data before performing any further analysis.  Unfortunately,
since incomplete sky coverage destroys the orthogonality of the
spherical harmonics, this procedure covertly removes part of the
contribution of the higher multipoles.  There are two ways to
compensate for this.  

The first option is to treat the monopole and dipole coefficients
($a_{00}$ and $a_{1m}$) as ``nuisance parameters,'' {\it i.e.},
quantities whose true values we neither know nor care about.\footnote{
A parameter may be a nuisance parameter at one time and an interesting
parameter at another.  For instance, if we want to estimate the
spectral index $n$, we should probably compute $L(\langle Q\rangle,n)$
and treat $\langle Q\rangle$ as a nuisance parameter.  At some
other time, though, 
we may think $\langle Q\rangle$ is an interesting thing to know.
}
In the
context of Bayesian analysis, the natural thing to do with nuisance
parameters it to marginalize over them.  Marginalizing over a nuisance
parameter $\zeta$ means replacing the likelihood $L$ with the
marginal likelihood
\begin{equation}
L_{\rm marg}=\int d\zeta L(\zeta) p(\zeta).
\end{equation}
Here $p(\zeta)$ is a prior probability density for $\zeta$, which
is usually taken to be constant.  By marginalizing over the data,
we are using a standard identity of probability theory,
\begin{equation}
p(x)=\int p(x\ |\ \zeta)p(\zeta)\,d\zeta,
\end{equation}
to remove all $\zeta$-dependence from the likelihood.

From a frequentist point of view, the natural way to get rid of a
nuisance parameter is to maximize with respect to it.  That is,
we replace $L$ with $\max_\zeta L(\zeta)$.  That way, a particular
model is ruled out only if it is ruled out for all possible values
of $\zeta$.

If we are performing some sort of data compression, then we have
a second option for dealing with the monopole and dipole.  We can
simply impose a constraint on our compression matrix $\bf A$, requiring
it to be insensitive to the unwanted multipoles.  This is in effect
the approach of G\'orski (1994): by orthogonalizing the spherical
harmonics, he makes his compression matrix insensitive to the monopole
and dipole.  This approach turns out to be mathematically equivalent
to marginalizing over the unwanted modes.

People frequently remove the quadrupole information from the COBE data
in the same way as the monopole and dipole, on the grounds that the
quadrupole is particularly susceptible to Galactic contamination.
It has also been known since the earliest days of COBE analysis that
the quadrupole is anomalously low (compared to the prediction of a
flat power spectrum normalized to the other multipoles).  From a statistical
point of view, this is a delicate situation: it is perfectly acceptable,
and even wise, to throw away data if there is a reasonable fear of
contamination, but throwing away data that is known {\it a priori}
to be discordant with favored theories is a major statistical
{\it faux pas}.  On balance, it is probably better to leave the
quadrupole information in in the interest of avoiding even the
possibility of biased editing of the data.

There is another argument in favor of retaining the quadrupole.
Even if the quadrupole is contaminated, it still contains useful information,
and so it may be unwise to throw it away entirely.  Since the quadrupole is
a root-mean-square quantity, any contaminant would tend to bias the quadrupole
up.  In fact, if a particular theory is ruled out because it predicts
too large a quadrupole,
hypothesizing an additional quadrupolar contaminant cannot
save that theory: as long as the contaminant 
is statistically independent of the cosmic signal, the
net result of hypothesizing a contaminant is  necessarily to {\em lower}
the likelihood of that theory.

\subsection{Results}

The main purpose of this section is to discuss data analysis
techniques, not results; however,
we will briefly present some results based on a Karhunen-Lo\`eve
analysis of the four-year COBE DMR data.  The reader is referred
to Bunn \& White (1996) for a more detailed discussion.

The data set used for this analysis consists of a weighted average
of the 53 and 90 GHz maps from the four-year DMR data.  The
maps are averaged with weights inversely proportional to the
noise variance, in order to minimize noise in the average
map.  (This is equivalent to performing a joint likelihood
analysis of the individual maps.)  We performed the Karhunen-Lo\`eve
analysis using a flat fiducial power spectrum $l(l+1)C_l=\mbox{const.}$, 
and we retained the 500 most significant modes.

\begin{figure}[t]
\centerline{\epsfig{file=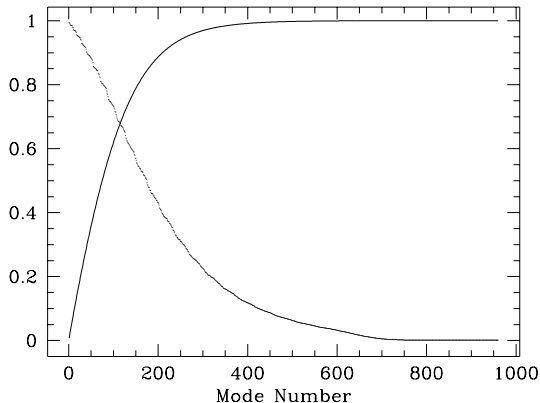,width=3in}}
\caption{The points show the eigenvalues $\lambda_a$ of the
four-year COBE data, sorted in decreasing order.  The
solid curve is the running sum of $\lambda_a^2$, normalized
to 1.}
\label{fig:eigvals}
\end{figure}

Figure \ref{fig:eigvals} shows the eigenvalues $\lambda_a$, together
with a running sum of the squares of the eigenvalues.  (Recall
that this sum is proportional to the rejection power $\gamma$.)
This plot indicates that modes beyond the first 500 do not
significantly increase our ability to discriminate among models.

\begin{figure}[t]
\centerline{\epsfig{file=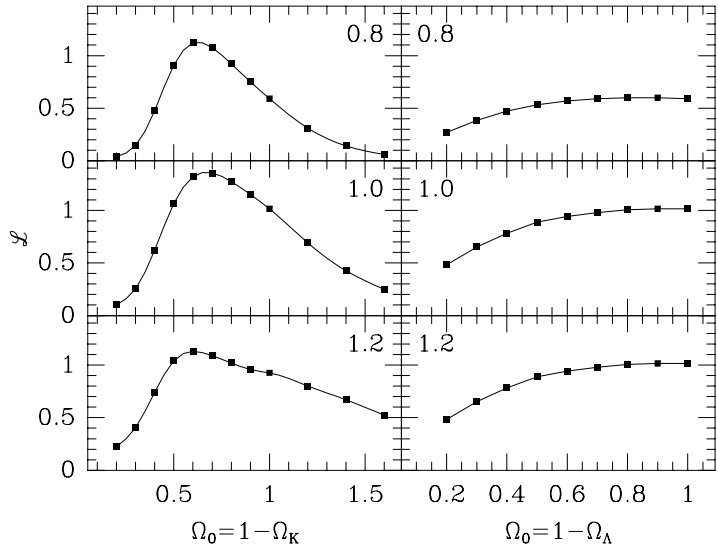,width=4in}}
\caption{Likelihood as a function of $\Omega_0$ for CDM models with
zero cosmological constant (left) and zero spatial curvature (right).  The
spectral index $n$ increases from 0.8 to 1.2 from top to bottom.
The likelihoods are normalized so that a flat spectrum has $L=1$.
See Bunn \& White (1996) for further details.}
\label{fig:likelycdm}
\end{figure}

\begin{figure}[t]
\centerline{\epsfig{file=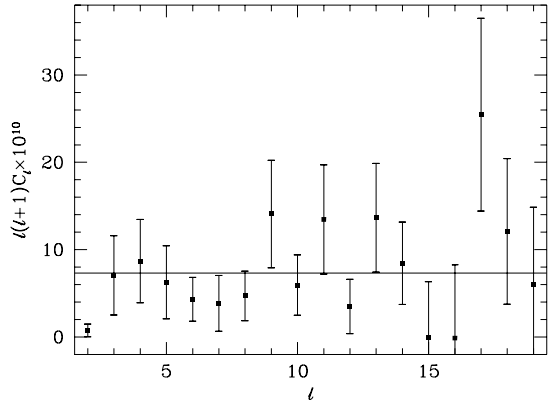,width=3in}}
\caption{The points represent the maximum-likelihood power spectrum,
obtained by letting all $C_l$'s between 2 and 19 vary freely.
A flat $\langle Q\rangle=19\,\mu\rm K$ power spectrum is plotted
for comparison.  The error bars are standard errors determined
by approximating the likelihood by a Gaussian near the peak.
Because the Gaussian approximation is poor, and because there are
significant correlations between the errors, these error bars
can be deceptive.  The small formal error on $C_2$ is particularly
misleading.  See Bunn \& White (1996) for further discussion.}
\label{fig:bestfit}
\end{figure}

Figure \ref{fig:likelycdm} shows the likelihood function for low-density
CDM models, both with and without a cosmological constant.  
Figure \ref{fig:bestfit} shows the maximum-likelihood power spectrum,
found by allowing each $C_l$ with $2\le l\le 19$ to vary independently.
The error bars shown in this figure are standard errors determined by
approximating the likelihood near the peak as a Gaussian.
The standard errors are then the square roots of the diagonal elements
of the covariance matrix of this Gaussian.  Error bars determined in this
way should be viewed with extreme caution.  First, the likelihood is not
very well approximated by a Gaussian: on the contrary, it is strongly
skew-positive at low $l$.  Second, these standard errors contain no
information about correlations between the errors.
These correlations are largest for pairs of modes whose $l$-values
differ by 2.  
(Coupling between modes with $\Delta l=1$ is weak because the data
have approximate reflection symmetry.)
The deceptively small error bar on the estimate of $C_2$
is largely due to the failure of the Gaussian approximation
for the likelihood, although the 15\% anticorrelation
between $C_2$ and $C_4$ also plays a role.

Finally, Table \ref{table:sigma8} shows values of the small-scale 
fluctuation amplitude $\sigma_8$ for various theoretical models.
The observational constraint is approximately $0.5\simless\sigma_8\simless
0.8$ ({\it e.g.}, Viana \& Liddle 1996).

\subsection{Wiener Filtering}

Until now, we have focused on attempts to estimate
the angular power spectrum $C_l$.  While this is the most useful
thing to do with a CMB data set, other complementary approaches
can be interesting in certain contexts.  For instance, we can
assume that we know the angular power spectrum and try to determine
the underlying cosmic signal from a noisy sky map.  That is,
we can attempt to {\em filter} a sky map, cleaning up the noise
and leaving the signal.  
The Wiener filter (Wiener 1949)
is optimal linear filter for this purpose, in the
sense of least squares.  The recent use of Wiener filtering in
astrophysics is largely due to Rybicki \& Press (1992),
and the filter has been applied to the
COBE data by Bunn, Hoffman, \& Silk (1996).

\begin{table}[t]
\begin{center}
\begin{tabular}{llllllll}
\hline
 & $\Omega_0$ & $\Omega_\Lambda$ & $\Omega_{\rm HDM}$ & $n$ & $h$
& $\Omega_{\rm B}h^2$ & $\sigma_8$ \\
\hline
standard CDM & 1.0 & 0.0 & 0.0 & 1.0 & 0.50 &  0.0125 &   1.22 \\
tilted CDM & 1.0 & 0.0 & 0.0 & 0.8 & 0.50 &  0.0250 &   0.72 \\
MDM & 1.0 & 0.0 & 0.2 & 1.0 & 0.50 &  0.0150 &   0.79 \\
$\Lambda$CDM & 0.4 & 0.6 & 0.0 & 1.0 & 0.65 &  0.0150 &   1.07 \\
Open CDM & 0.4 & 0.0 & 0.0 & 1.0 & 0.65 &  0.0150 &   0.64 \\
Low $h$ CDM &  1.0 & 0.0 & 0.0 & 1.0 & 0.35 & 0.0150 &   0.74\\
\hline
\end{tabular}
\end{center}
\caption{The predicted fluctuation amplitude on scales of $8\,h^{-1}\rm
Mpc$ for various CDM-like models.  MDM is a ``mixed dark matter''
model.  All normalizations are from the four-year COBE DMR data.
See Bunn \& White (1996) for further details.}
\label{table:sigma8}
\end{table}

Suppose we have a data
vector $\vec d$ containing signal and noise.  We want to apply
a linear filter $\bf F$ so that $\vec y\equiv {\bf F}\vec d$
approximates the true cosmic signal $\Delta T/T$ in such a way that
the mean-square deviation,
\begin{equation}
\left<\left(y_i-{\Delta T\over T}(\hat{\bf r}_i)\right)^2\right>,
\end{equation}
is as small as possible.  The solution to this optimization problem
is the Wiener filter,
\begin{equation}
{\bf F}={\bf M}_{\rm signal}{\bf M}^{-1},
\end{equation}
where $\bf M$ is as usual the data covariance matrix and ${\bf M}_{\rm signal}$
is the signal contribution to $\bf M$.

Under the assumption of Gaussian statistics, the Wiener-filtered
data is also the maximum-likelihood estimator of $\Delta T/T$ at
each point.  Note that in regions of very high noise, where we
have little information, the Wiener filter returns values near zero,
because this is the most likely {\it a priori}
value of a zero-mean Gaussian.

\begin{figure}[t]
\centerline{\epsfig{file=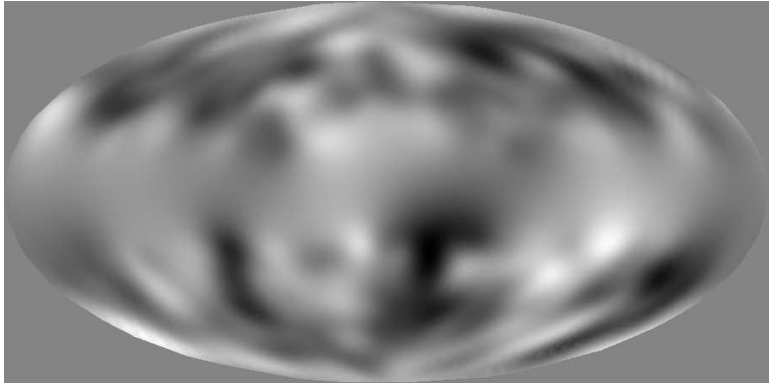,width=4in}}
\caption{A sky map, in Aitoff projection, of the Wiener-filtered
four-year DMR data.  The relative lack of structure near the Galactic
plane is due to the fact that no data from that region were used.}
\label{fig:wiener}
\end{figure}

Figure \ref{fig:wiener} shows a Wiener-filtered COBE sky map.  Although
the signal-to-noise ratio in the raw pixel maps is typically less
than one per pixel, the largest-amplitude features in the filtered
map are significant at the five sigma level per pixel.

One of the main uses of the filtered maps is in making predictions
for other experiments.  Assuming Gaussian statistics, the full
error covariance matrix of the Wiener-filtered map is known, and
so we can produce maps with known uncertainties of a region of the
sky.  For predictions of the CMB sky as it should be seen by the
Tenerife experiment, see Bunn, Hoffman, \& Silk (1996).

\section{Summary}
\label{sec:summary}

The main lesson to be learned from this entire institute is that this
is an exciting time in CMB research.  The existing data are already
telling us vast amounts about cosmology, and in the next few years the
data should continue to improve dramatically.  The high quality
of present and future anisotropy observations presents us with some challenges.
We must understand our theoretical models well enough to make accurate
predictions, and we must develop statistical tools that enable us
to determine which predictions are consistent with the data.
Both of these challenges are currently being met with ever-increasing
success.

The tools for making accurate predictions, at least in linear
models like CDM, are by now quite well developed.  Furthermore,
in recent years analytic and semianalytic approximations
have dramatically improved
our understanding of the basic physical principles involved
in anisotropy formation.

The problem of data analysis is also much better understood
today than it was five years ago (before there were any actual
detections to analyze).  However, it is important to remember that
analysis of future data sets will present challenges that make the
COBE analysis look easy.  
When sky maps contain a million pixels
instead of a few thousand, data compression will be absolutely
essential.  It is already time to start thinking about this 
difficult problem.

In addition, future experiments with higher resolution than COBE
will be more susceptible to foreground contamination.  In the
case of COBE, it is believed that simply excising points
too close to the Galactic plane is sufficient to remove most
of the foreground contamination; for future high-sensitivity
degree-scale experiments, more sophisticated methods will be
necessary.

We have seen that CDM-like theoretical models predict that vast
am\-ounts of information are encoded in the CMB anisotropy power
spectrum.  There is a very real hope that the CMB will give us
accurate values for all sorts of cosmological parameters.  But
even if the information is there, we will have to do a lot
of work to wrest it from the data.

\section{Acknowledgments}

The first half of these lectures is based largely on the work of Wayne
Hu and Naoshi Sugiyama.  Much of the later material is based on 
work I performed in collaboration with Douglas Scott, Joseph Silk,
Max Tegmark, and Martin White.  I would like to thank all of these
people for many helpful discussions.  In addition, Wayne, Max, and Martin
made some of the figures.  Finally, I would like to thank the organizers
of the meeting for their hard work and hospitality.


\begin{thebibliography}{88}

\bibitem{}
Banday, A. \etal~1996, preprint (astro-ph/9601065).

\bibitem{}
Bennett, C.L. \etal~1992, \apjl, 396, 6.

\bibitem{}
Bennett, C.L. \etal~1996, \apjl, 464, 1.

\bibitem{}
Berger, J.O. 1985, {\it Statistical Decision Theory and Bayesian Analysis}
(Springer-Verlag).

\bibitem{}
Blandford, R.D. \& Narayan, R.~1992, Ann. Rev. Astron. Astrophys., 30, 311.

\bibitem{}
Bond, J.R. 1994, in {\it Proceedings of the IUCAA Dedication Ceremonies},
(ed. T. Padmanabhan; John Wiley \& Sons).

\bibitem{}
Bond, J.R. 1995, Phys. Rev. Lett., 74, 4369.

\bibitem{}
Bond, J.R. 1996, in {\it Cosmology and Large-Scale Structure}
(1994 Les Houches summer school; ed., R. Schaefer; Elsevier, in press.)

\bibitem{}
Bond, J.R. \& Efstathiou, G. 1984, \apjl, 285, 45.

\bibitem{}
Bond, J.R. \& Efstathiou, G. 1987, \mnras, 226, 655.

\bibitem{}
Bunn, E.F. 1995, {\it Statistical Analysis of Cosmic Microwave 
Background Anisotropy}, Ph.D. thesis, Physics Department, U.C. Berkeley \\
(ftp://pac2.berkeley.edu/pub/bunn/thesis).

\bibitem{}
Bunn, E.F., Hoffman, Y., \& Silk, J.~1996, \apj, 464, 1.

\bibitem{} 
Bunn, E.F., Liddle, A., \& White, M. 1996, preprint (astro-ph/9607038).

\bibitem{} 
Bunn, E.F., Scott, D., \& White, M. 1995, \apjl, 441, 9.

\bibitem{}
Bunn, E.F. \& Sugiyama, N. 1995, \apj, 446, 49.

\bibitem{} Bunn, E.F. \& White, M. 1996, preprint
(astro-ph/9607060).

\bibitem{} 
Bunn, E.F., White, M., Srednicki, M., \& Scott, D. 1994, \apj, 429, 1.

\bibitem{}
Dodelson, S. \& Jubas, J. 1995, Ap.J., 439, 503.

\bibitem{}
Doroshkevich, A.G., Zel'dovich, Ya.B., \& Sunyaev, R. 1978, 
Sov. Astron., 22, 523.

\bibitem{}
G\'orski, K. 1994, \apjl, 430, 85.

\bibitem{}
G\'orski, K. \etal~1994, \apjl, 430, 89.

\bibitem{}
G\'orski, K. \etal~1996, \apjl, 464, 11.

\bibitem{}
Gull, S.F. \& Daniell, G.J. 1978, Nature, 272, 686.

\bibitem{}
Gunn, J.M. \& Peterson, B.A. 1965, \apj, 142, 1663.

\bibitem{} 
Hinshaw, C. \etal~1996, \apjl, 464, 17.

\bibitem{}
Hu. W. 1995, {\it Wandering in the Background: A Cosmic
Background Explorer},
Ph.D. Thesis, Physics Department, U.C. Berkeley \\
(ftp://pac2.berkeley.edu/pub/hu/thesis).

\bibitem{}
Hu, W. 1996, in
{\it The Universe at High-z, Large-Scale Structure and the
Cosmic Microwave Background}, eds. E. Martinez-Gonzalez \&
J.-L. Sanz (Springer-Verlag) (astro-ph/9511130).

\bibitem{}
Hu, W., Scott, D., \& Silk, J. 1994, Phys. Rev. D, 49, 648.

\bibitem{}
Hu, W., Scott, D., Sugiyama, N. \& White, M. 1995, Phys. Rev. D., 52,
5498.

\bibitem{}
Hu, W. \& Sugiyama, N. 1994, in {\it CWR CMB Workshop: Two
Years After COBE}, eds. L. Krauss, P. Kernan (World Scientific, Singapore,
p. 188).

\bibitem{} 
Hu. W. \& Sugiyama, N. 1995a, Phys. Rev. D, 51, 2599.

\bibitem{} 
Hu, W. \& Sugiyama, N. 1995b, \apj, 444, 489.

\bibitem{} 
Hu. W. \& Sugiyama, N. 1996, preprint (astro-ph/9510117).

\bibitem{}
Hu, W., Sugiyama, N., \& Silk, J. 1996, Nature, in press (astro-ph/9604166;
see also http://www.sns.ias.edu/\lower0.05in\hbox{\char'176}\kern-0.02in
whu/physics/physics.html). 

\bibitem{}
Hu, W. \& White, M. 1996, in {\it Proceedings of the XXXIst Moriond
Meeting, Microwave Background Anisotropies}, in press.

\bibitem{}
Jackson, J.D. 1975, {\it Classical Electrodynamics} (Wiley).

\bibitem{}
Jungman, G., Kamionkowski, M., Kosowsky, A., \& Spergel, D.~1995, 
preprint, astro-ph/9512139.

\bibitem{}
Karhunen, K. 1947, {\it \"Uber Lineare Methoden in der 
Wahrscheinlichkeitsrechnung} (Kirjapaino oy. Sana, Helsinki).

\bibitem{}
Kosowsky, A., Kamionkowski, M., Jungman, G., \& Spergel, D.N. 1996
preprint (astro-ph/9605147).

\bibitem{}
Lineweaver, C.H. \etal~1994, \apj, 436, 452.

\bibitem{}
Netterfield, C.B., Devlin, M.J., Jarosik, N., Page, L., \& Wollack, E.J.
1996, preprint (astro-ph/9601197).

\bibitem{}
Ostriker, J. \& Vishniac, E. 1986, \apj, 306, 51.

\bibitem{}
Peebles, P.J.E. 1987, Nature, 327, 210.

\bibitem{}
Peebles, P.J.E. \& Yu, J.T. 1970, \apj, 162, 815.

\bibitem{}
Press, W.H. 1996, in {\it Unsolved Problems in Astrophysics}
(ed. J.P. Ostriker, Princeton University Press, in press, astro-ph/9604126).

\bibitem{}
Rees, M. \& Sciama, D. 1968, Nature, 519, 611.

\bibitem{}
Rice, J.A. 1995, {\it Mathematical Statistics and Data Analysis}
(Duxbury).

\bibitem{}
Rybicki, G. \& Press, W. 1992, \apjl, 432, 75.

\bibitem{}
Sachs, R.K. \& Wolfe, A.M. 1967, \apj, 147, 73.

\bibitem{}
Scaramella, R. \& Vittorio, N. 1993, \mnras, 263, 17.

\bibitem{}
Scott, D., Silk, J. \& White, M. 1995, Science, 268, 829.

\bibitem{} 
Seljak, U. 1994, \apjl, 435, 87.

\bibitem{}
Seljak, U. 1996a, \apj, 460, 549. 

\bibitem{}
Seljak, U. 1996b, \apj, 463, 1.. 

\bibitem{}
Seljak, U. \& Bertschinger, E. 1994, \apjl, 417, 9.

\bibitem{}
Seljak, U. \& Zaldarriaga, M. 1996, preprint (astro-ph/9603033).  

\bibitem{}
Silk, J. 1967, Nature, 215, 1155.

\bibitem{}
Silk, J. 1968, \apjl, 151, 459.

\bibitem{}
Silk, J. 1982, Acta Cosmologica, 11, 75.

\bibitem{}
Smoot, G. \etal~1992, \apjl, 396, 1.

\bibitem{}
Sunyaev, R.A. 1977, Sov. Astron. Lett., 3, 491.

\bibitem{}
Sunyaev, R.A. \& Zel'dovich, Ya.B. 1970, Astrophys. Space Sci., 7, 3.

\bibitem{}
Sunyaev, R.A. \& Zel'dovich, Ya.B. 1972, Comm. Astrophys. Space Phys., 4, 73.

\bibitem{} 
Tegmark, M. 1995, \apj, 455, 429.

\bibitem{} 
Tegmark, M. 1996a, \mnras, 280, 299.

\bibitem{} 
Tegmark, M. 1996b, \apjl, 464, 35.

\bibitem{}
Tegmark, M. 1996c, 
to appear in Proc. Enrico Fermi, Course CXXXII, Varenna (astro-ph/9511148).

\bibitem{}
Tegmark, M. \& Bunn, E.F. 1995, \apj, 451, 1.

\bibitem{}
Tegmark, M. \& Efstathiou, G. 1996, preprint (astro-ph/9507009).

\bibitem{} 
Tegmark, M., Taylor, A., \& Heavens, A. 1996, 
preprint (astro-ph/9603021).

\bibitem{}
Viana, P.T.P. \& Liddle, A. 1996, \mnras, in press (astro-ph/9511007).

\bibitem{}
Vittorio, N. \& Silk, J. 1984, \apjl, 285, 39.

\bibitem{}
Vogeley, M.S. \& Szalay, A.S. 1996, \apj, 465, 34.

\bibitem{}
White, M. \& Bunn, E.F. 1995, \apj, 450, 477.

\bibitem{}
White, M., Scott, D., \& Silk, J. 1994, Ann. Rev. Astron. Astrophys., 32, 319.

\bibitem{}
White, M. \& Srednicki, M. 1995, \apj, 443, 6.

\bibitem{}
Wiener, N. 1949, {\it Extrapolation and Smoothing of Stationary Time
Series} (Wiley).

\bibitem{}
Wilson, M. 1983, \apj, 273, 2.

\bibitem{}
Wilson, M. \& Silk, J. 1981, \apj, 243, 14.

\bibitem{}
Wright, E.L. \etal~1992, \apjl, 396, 11.

\bibitem{}
Wright, E.L. \etal~1994a, \apj, 420, 1.

\bibitem{}
Wright, E.L. \etal~1994b, \apj, 436, 443.

\bibitem{}
Zel'dovich, Ya.B. \& Sunyaev, R., Astrophys. Space Sci., 4, 301 (1969).

\end{thebibliography}
\end{document}